# Spin Hall magnetoresistance in a low-dimensional magnetic insulator

Saül Vélez[1,2,*], Vitaly N. Golovach[3,4,5,**], Juan M. Gomez-Perez[1], Cong Tinh Bui[6,7], F. Rivadulla[6], Luis E. Hueso[1,5], F. Sebastian Bergeret[3,4,†], and Fèlix Casanova[1,5,††]

[1]*CIC nanoGUNE, 20018 Donostia-San Sebastian, Basque Country, Spain*
[2]*Department of Materials, ETH Zürich, 8093 Zürich, Switzerland*
[3]*Centro de Física de Materiales (CFM-MPC), Centro Mixto CSIC-UPV/EHU, 20018 Donostia-San Sebastian, Basque Country, Spain*
[4]*Donostia International Physics Center (DIPC), 20018 Donostia-San Sebastian, Basque Country, Spain*
[5]*IKERBASQUE, Basque Foundation for Science, 48013 Bilbao, Basque Country, Spain*
[6]*Centro Singular de Investigación en Química Biolóxica e Materiales Moleculares (CiQUS), and Departamento de Química-Física, Universidad de Santiago de Compostela, 15782 Santiago de Compostela, Spain*
[7]*Department of Electrical and Electronic Engineering, Tokyo Institute of Technology, 2-12-1 Ookayama, Meguro, Tokyo 152-0033, Japan*

[*] saul.velez@mat.ethz.ch
[**] vitaly.golovach@ehu.eus
[†] sebastian_bergeret@ehu.eus
[††] f.casanova@nanogune.eu

**Abstract**

We observe an unusual behavior of the spin Hall magnetoresistance (SMR) in Pt deposited on a tensile-strained LaCoO$_3$ (LCO) thin film, which is a ferromagnetic insulator with the Curie temperature $T_c$=85K. The SMR displays a strong magnetic-field dependence below $T_c$, with the SMR amplitude continuing to increase (linearly) with increasing the field far beyond the saturation value of the ferromagnet. The SMR amplitude decreases gradually with raising the temperature across $T_c$ and remains measurable even above $T_c$. Moreover, no hysteresis is observed in the field dependence of the SMR. These results indicate that a novel low-dimensional magnetic system forms on the surface of LCO and that the LCO/Pt interface decouples magnetically from the rest of the LCO thin film. To explain the experiment, we revisit the derivation of the SMR corrections and relate the spin-mixing conductances to the microscopic quantities describing the magnetism at the interface. Our results can be used as a technique to probe quantum magnetism on the surface of a magnetic insulator.

***Introduction.**–*Magnetoresistance has been key for understanding spin-dependent transport in solids [1]. In the last years, new magnetoresistance phenomena were discovered in thin ferromagnetic/normal metal(FM/NM)-based heterostructures [2–18], which originate from the interplay of the spin currents generated in the heterostructure (*via* the spin Hall effect [19–22] or the Rashba-Edelstein effect [23,24]) with the magnetic moments of the FM layer. Among many applications, these magnetoresistance effects have been used for quantifying spin transport properties such as the spin diffusion length $\lambda$ and the spin Hall angle $\theta_{SH}$ of different NM layers, or the spin-mixing conductance $G_{\uparrow\downarrow}$ of FM/NM interfaces. More interestingly, unlike other surface-sensitive techniques that suffer from a bulk contribution due to a finite penetration depth, the spin Hall magnetoresistance (SMR) [4–11] uses the spin accumulation at interfaces for sensing the magnetic properties of the very first atomic layer of magnetic insulators (MIs) [25,26]. For instance, SMR has been employed for probing the surface of complex magnetic systems such as ferrimagnetic spinel oxides [11,27], spin-spiral multiferroics [28,29], canted ferrimagnets [30], $Y_3Fe_5O_{12}$/antiferromagnetic (YIG/AFM) bilayers [31,32], and synthetic AFMs [33].

$LaCoO_3$ (LCO) presents an intriguing magnetic behavior, which has been studied for decades and is still under debate [34–49]. Bulk LCO is a diamagnetic insulator at low temperature, owing to the low-spin (LS) configuration of $Co^{3+}$. The relatively small crystal-field splitting of the $Co^{3+}$ 3d-shell results in an increasing population of high-spin (HS) $Co^{3+}$ with temperature, reaching 1:1 (LS:HS) above ~150K. The close proximity between crystal-field splitting and exchange energy makes the magnetic properties of LCO particularly susceptible to small changes in inter-ionic distances and coordination. For this reason, tensile-strained LCO thin films grown on particular substrates [such as $SrTiO_3$ (STO)] exhibit FM order at low temperatures [42–49]. However, the magnetic properties of the surface of these films –where the crystal-field symmetry is lowered because of a different stoichiometry at the surface– have not been addressed yet.

In this letter, we take the first steps towards understanding the magnetic behavior of the surface of strained LCO films by performing magnetoresistance measurements in STO/LCO/Pt. We find that SMR depends strongly on the magnetic field at all temperatures, both above and below the Curie temperature ($T_c$) of the film, and more strikingly, no hysteresis in the magnetoresistance is observed. These observations clearly show that the surface magnetism of the LCO film is radically different from its bulk counterpart. We support our measurements with a theoretical model that extends the known expressions for SMR [7,50] and HMR [51,52] in MI/NM bilayers for an arbitrary magnetic ordering (para-, ferri-, ferro-, antiferro-magnet) of the localized magnetic moments at the MI/NM interface. We provide expressions for $G_{\uparrow\downarrow}=G_r+iG_i$ [25,53] and the effective spin conductance $G_s$ [54,55] in terms of surface spin-correlators. The experimental data evidence that the surface of LCO behave as a low-dimensional Heisenberg FM.

*Experimental details.*–Growth of epitaxial LCO thin films via polymer-assisted deposition on (001) STO substrates, as well as their structural, electrical, and magnetic characterization, is described in Ref. [46]. The LCO films exhibit a tetragonal distortion, which induces FM ordering below $T_c$~85K and with a coercive field below 1T at 10K, in agreement to other reports [43–45,47,56]. The films exhibit low surface roughness (<1nm) and are insulating [46]. Pt Hall bar structures (width $W$=100μm, length $L$=800μm and thickness $d_N$=7nm) were patterned on top of the LCO films *via* e-beam lithography, sputtering deposition of Pt, and lift-off. Two samples with different LCO thickness, 12nm and 19nm, were prepared and studied, showing similar results. Below, we present data for the 19-nm-thick LCO film. Magnetotransport measurements were performed between 10 and 300K in a liquid-He cryostat that allows applying magnetic fields $H$ of up to 9T and rotating the sample by 360º.

*Longitudinal magnetoresistance in LCO/Pt.*–Figures 1(a)-1(f) show the longitudinal angular-dependent magnetoresistance (ADMR) in LCO/Pt at 200 and 70K and for $H$=9T in the three relevant **H**-rotation planes $\alpha,\beta,\gamma$ (see sketches). We can see a clear ADMR with a $\cos^2$ modulation in $\alpha$ and $\beta$, and almost no variation in $\gamma$. These angular dependences are in agreement with the ones expected for spin-related magnetoresistances, such as SMR and HMR [5,52]. Surprisingly, this angular symmetry is not only observed below $T_c$ of LCO [Figs. 1(d)-(f)] but also above, *i.e.*, when the LCO film is in the paramagnetic (PM) state [Figs. 1(a)-(c)].

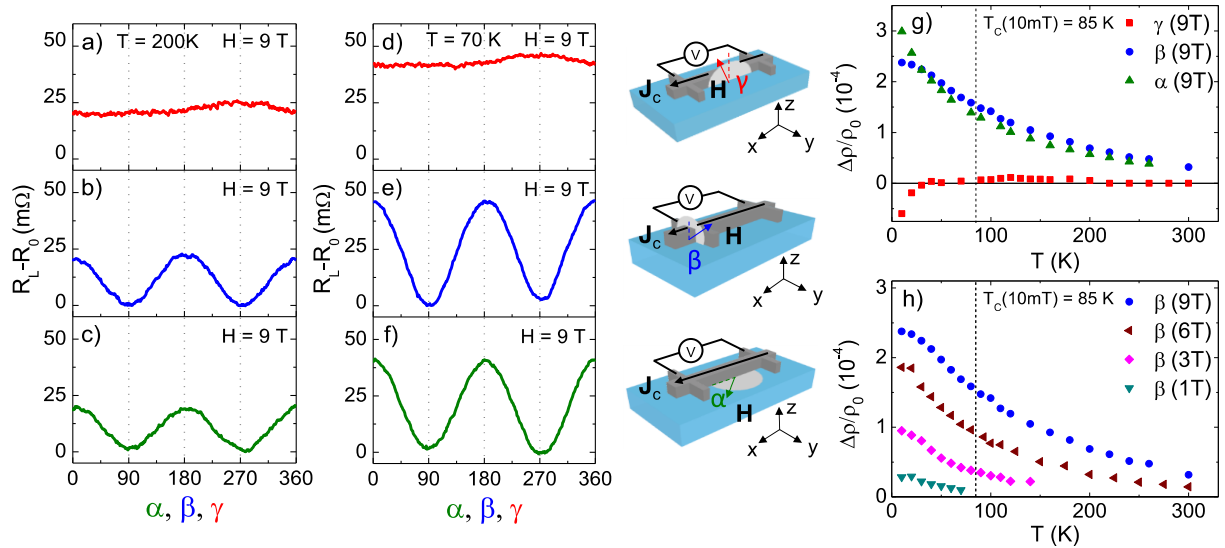

FIG. 1. Longitudinal ADMR measurements performed in LCO(19nm)/Pt(7nm) at (a)-(c) 200K and (d)-(f) 70K for $H$=9T in the $\alpha$–, $\beta$–, and $\gamma$–rotation planes. The sketches indicate the definition of the angles, the axes, and the measurement configuration. $R_0$ is taken as $R_L(\alpha,\beta$=90º). (g) Temperature dependence of the normalized ADMR amplitude, $\Delta\rho/\rho_0 \cong (R_L(0º)-R_L(90º))/R_L(90º)$, measured at 9T in the $\alpha$–, $\beta$–, and $\gamma$–planes (the data is fitted to a $\cos^2$ dependence). $\rho_0$ is the Drude resistivity. (h) Temperature dependence of $\Delta\rho/\rho_0$ measured in the $\beta$–plane for different $H$ values. Vertical dashed lines in (g) and (h) indicate $T_c$ of LCO at 10mT [26].

Figure 1(g) shows the temperature dependence of the ADMR amplitude measured at 9T in the $\alpha$–, $\beta$–, and $\gamma$–planes. The amplitude is roughly the same in $\alpha$ and $\beta$ and decays monotonously with temperature, whereas it is negligibly small in $\gamma$, except for very low temperatures. The sign of the ADMR in $\gamma$ and the increase in $\alpha$ below ~20K suggest the emergence of magnetic proximity effect (MPE) at the LCO/Pt interface at low temperatures. The MPE could be at the origin of the unusual temperature-dependence of the Hall resistance reported in this system [56], an unconventional behavior that is also observed in our sample [26].

Figure 1(h) shows that the ADMR amplitude depends strongly on $H$ at all temperatures. However, since the magnetization of the LCO saturates above $H$~1T in the FM phase [46], no variation of the ADMR amplitude is expected for H>1T [7]. Besides, our measurements show a smooth change of the ADMR amplitude across the FM-PM transition, with a significant magnetoresistance measured even far above $T_c$ [see Figs. 1(g)-1(h)]. In contrast, a sudden drop in the magnetoresistance is expected to take place when the film becomes PM. All these observations indicate that the magnetic response of the surface of the LCO film is decoupled from its bulk.

For a better understanding of the origin of the magnetoresistance we measure the longitudinal field-dependent magnetoresistance (FDMR) along the three main axes of the sample and for different temperatures. Figures 2(a)-2(b) show representative FDMR curves obtained far above (200K) and below (70K) $T_c$. The data indicate that the magnetoresistance in each regime should have different origins. For $T>>T_c$, the FDMR along the $y$-direction (i.e., the direction of the polarization of the spin accumulation in Pt) is rather constant, whereas equal parabolic-like FDMR curves are obtained in the $x$- and $z$-directions. This behavior is characteristic of the HMR effect in thin films with strong spin-orbit coupling [52].

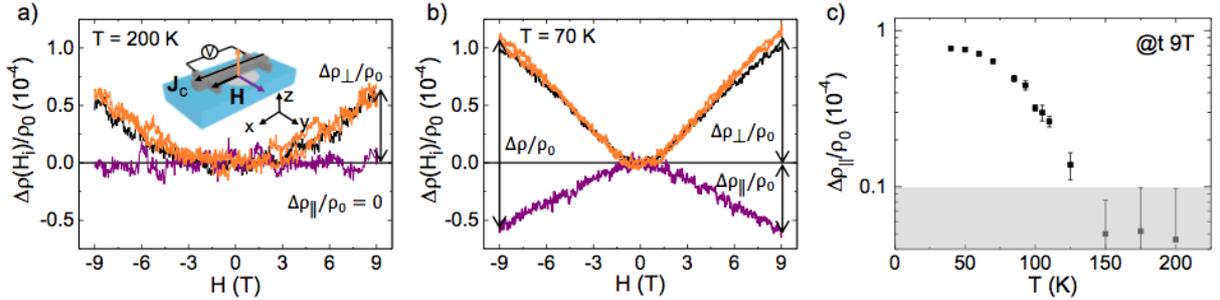

FIG. 2. Normalized FDMR measurements, $\Delta\rho(H_i)/\rho_0 \cong (R_L(H_i)-R_L(0))/R_L(0)$ ($H_i$ indicates that the magnetic field is applied along the $i$-direction), performed in LCO(19nm)/Pt(7nm) at (a) 200K and (b) 70K along the three main sample axes. The sketch in (a) indicates the definition of the axes, color code of the magnetic field direction, and measurement configuration. (c) Temperature dependence of $\frac{\Delta\rho_\parallel}{\rho_0}$ at 9T [see (b) for its definition]. The shaded region indicates the noise floor.

For $T<T_c$, the three FDMR curves lie on the same resistance value at $H$=0 [Fig. 2(b)]. When the magnetic field is increased, a magnetoresistance symmetric with $H$ develops, having equal positive amplitudes in $x$- and $z$-directions, and a smaller and negative amplitude in the $y$-direction. Moreover, no hysteresis is observed between the trace and retrace curves. These observations are in sharp contrast with those found in other magnetic systems, such as YIG [5,8–10,52] and CoFe$_2$O$_4$ [27], where the FM order results in hysteretic FDMR curves and different resistance states around $H$=0 for different field directions. Therefore, the FDMR measurements shown in Fig. 2(b) do not reflect the bulk FM properties of the LCO film and support the idea that the surface is magnetically decoupled.

From Figure 2b, we can see that the total magnetoresistance $\frac{\Delta\rho}{\rho_0} = \frac{\rho(H_z)-\rho(H_y)}{\rho_0} \cong \frac{\rho(H_x)-\rho(H_y)}{\rho_0}$ (i.e., the ADMR amplitude, Fig. 1) has two contributions: $\frac{\Delta\rho_\perp}{\rho_0} = \frac{\rho(H_z)-\rho(0)}{\rho_0} \cong \frac{\rho(H_x)-\rho(0)}{\rho_0}$ and $\frac{\Delta\rho_\parallel}{\rho_0} = \frac{\rho(0)-\rho(H_y)}{\rho_0}$. In the high temperature regime (Fig. 2a), $\frac{\Delta\rho}{\rho_0} \cong \frac{\Delta\rho_\perp}{\rho_0}$ and $\frac{\Delta\rho_\parallel}{\rho_0} \approx 0$, which is consistent with HMR [52]. At low temperatures, however, we observe a finite contribution from $\frac{\Delta\rho_\parallel}{\rho_0}$ (Fig. 2b). As we will demonstrate below, this contribution emerges from the magnetic response of the LCO/Pt interface. Figure 2(c) shows the temperature dependence of $\frac{\Delta\rho_\parallel}{\rho_0}$ at $H_y$=9T. This contribution is larger at low temperatures, decreases monotonically with increasing temperature, and drops below our resolution limit at $T$~125-150K, far above $T_c$. Therefore, one cannot attribute the suppression of $\frac{\Delta\rho_\parallel}{\rho_0}$ merely to the FM-PM transition of the bulk LCO film. This temperature dependence is yet another strong evidence that the magnetic response of the LCO/Pt interface must be decoupled from the bulk of the LCO film.

A different magnetic response for the surface Co atoms in LCO films is likely expected because the octahedral coordination of the $Co^{3+}$ atoms that lead to the FM order [48] is broken at the surface. What is surprising, however, is that the surface and the bulk of the film are magnetically decoupled. In analogy to what has been observed in other oxide layers, such as in $La_{0.7}Sr_{0.3}MnO_3$ [57], we speculate that the decoupling of the surface Co atoms in LCO is induced by a ferrodistortive Co ion off-centering, which can be promoted by the itinerant electrons in the Pt layer by pulling off the Co surface ions. Additionally, oxygen vacancies at the surface, a common effect observed in $Co^{3+}$ oxides [58,59], would also imply surface reconstruction, which would favour a different magnetic behaviour for the surface and possibly decoupling. For a more detailed discussion, see Ref. [26].

*Modeling.*–All the results presented above indicate that the LCO surface exhibits a PM-like behaviour. Spin-dependent phenomena, including spin pumping and spin Seebeck effect, have been recently reported using PM materials and ascribed to the presence of short-range FM correlations [60,61]. However, current existing theories on SMR only consider ferromagnetic ordering in the MI [7,50]. Here, we present a generalized theoretical model that describes the magnetoresistance in MI/NM bilayers including both SMR [7,50] and HMR [51,52] effects, as well as allows different magnetic orderings in the MI by describing microscopically the spin transport across the interface.

We model the MI/NM interface (*x-y* plane) as an ensemble of localized moments with spin **S**. These moments interact with the conduction electrons *via* an exchange term $\mathcal{H} = -J_{sd}\sum_j \boldsymbol{S}_j \cdot \boldsymbol{s}(\boldsymbol{r}_j)$, where $J_{sd}$ is the s-d exchange coupling and $\boldsymbol{s}(\boldsymbol{r}_j)$ is the spin density of the itinerant electrons at the position of the local moment $\boldsymbol{S}_j$. The spin current at the MI/NM interface is given by [26]:

$$-e\boldsymbol{J}_{s,z} = G_s\boldsymbol{\mu}_s + G_r\boldsymbol{n}\times[\boldsymbol{n}\times\boldsymbol{\mu}_s] + G_i\boldsymbol{n}\times\boldsymbol{\mu}_s, \qquad (1)$$

where $e>0$ is the elementary charge, $J_{s,z}$ is the spin current flowing in $z$-direction, $\boldsymbol{\mu_s}$ the vector spin accumulation, and $\boldsymbol{n}$ a unit vector in the direction of the applied magnetic field $\boldsymbol{B}=\mu\boldsymbol{H}$, with $\mu \approx \mu_0$ the magnetic permeability of the NM layer. The parameters $G_{r,i,s}$ are obtained in the Born approximation and are defined in terms of spin averages [26]:

$$G_r = \frac{\pi(\nu J_{sd})^2 e^2 n_s}{\hbar}\left(\langle S_\parallel^2\rangle - \frac{\langle S_\perp^2\rangle}{2}\right),$$
$$G_i = -\frac{\nu J_{sd} e^2 n_s}{\hbar}\langle S_\parallel\rangle,$$
$$G_s = -\frac{\pi(\nu J_{sd})^2 e^2 n_s}{\hbar}\langle S_\perp^2\rangle, \qquad (2)$$

where $\nu$ is the density of electronic states per spin species in the NM, $n_s$ is the surface density of localized magnetic moments at the MI/NM interface, and $\hbar$ the reduced Planck constant. $S_{\parallel,\perp}$ are the components of the spin operator parallel and perpendicular to $\boldsymbol{H}$. The dependence of the averages $\langle S_{\parallel,\perp}^2\rangle$ and $\langle S_\parallel\rangle$ with $\boldsymbol{H}$ and $T$ are determined by the type of magnetic order at the interface and, for instance, can be computed analytically for a PM.

In order to compute the magnetoresistance, we solve the spin diffusion equation in the NM layer subjected to the boundary condition imposed by Eq. (1) at the MI/NM interface and vanishing spin current at the interface with vacuum [7,50,52], and obtain the general expression for the longitudinal resistivity in leading order of $\theta_{SH}$: $\rho_L = \frac{1}{\sigma_0} + \Delta\rho_0 + \Delta\rho_1(1-n_y^2)$. Here $n_y$ is the $y$-component of $\boldsymbol{n}$, $\sigma_0 = 1/\rho_0$ is the conductivity of the NM layer, and the corrections $\Delta\rho_{0,1}$ are given by

$$\Delta\rho_0 = \frac{2\theta_{SH}^2}{\sigma_0}\left(1 - \frac{\lambda}{d_N}\frac{\tanh\left(\frac{d_N}{2\lambda}\right)-\frac{G_s\lambda}{\sigma_0}}{1-2\frac{G_s\lambda}{\sigma_0}\coth\frac{d_N}{\lambda}}\right),$$
$$\Delta\rho_1 = \frac{2\theta_{SH}^2}{\sigma_0}\left\{\frac{\lambda}{d_N}\frac{\tanh\left(\frac{d_N}{2\lambda}\right)-\frac{G_s\lambda}{\sigma_0}}{1-2\frac{G_s\lambda}{\sigma_0}\coth\frac{d_N}{\lambda}} - \Re\left[\frac{\Lambda}{d_N}\frac{\tanh\left(\frac{d_N}{2\lambda}\right)+\frac{G\Lambda}{\sigma_0}}{1+2\frac{G\Lambda}{\sigma_0}\coth\frac{d_N}{\Lambda}}\right]\right\}, \qquad (3)$$

where $\frac{1}{\Lambda} = \sqrt{\frac{1}{\lambda^2} + i\frac{1}{\lambda_m^2}}$ with $\lambda_m = \sqrt{\frac{D\hbar}{g\mu_B|B|}}$, $D$ the diffusion coefficient, $g$ the gyromagnetic factor, $\mu_B$ the Bohr magneton, and $G = (G_r - G_s + iG_i)$.

Equations (3) generalize the magnetoresistance in MI/NM bilayers in two ways: (i) They include the effective spin conductance $G_s$, so far omitted in SMR, which accounts for the fact that not all magnetic moments at the MI/NM interface may align in the field direction and hence the correlation $\langle S_\perp^2\rangle$ becomes finite. In the limit $G_s \to 0$, these equations recover both the previously reported SMR and HMR corrections [7,52], merged in a single analytical expression. (ii) They contain implicitly information about the magnetic response of the MI through the temperature and field dependence of the spin conductances defined in Eqs. (2). Note that $G_s$ is related to the ability of the spin accumulation to emit magnons in the MI and enters in Eq. (3) correcting $G_r$ as $G_r$-$G_s$, which stands for the effective spin-relaxation at the interface. The implications of Eqs. (2)-(3) are multiple. For instance, they can be used for understanding the temperature dependence of the SMR [62–64] or accurately describing the field-dependence of the magnetoresistance for non-collinear or non-saturated magnets. Particularly interesting is that Eqs. (2)-(3) are also valid across magnetic phase transitions [28,32,65,66]. Furthermore, note that Eq. (2) is not restricted to SMR and can also be applied to describe other spin transport phenomena involving interfaces such as electrical magnon excitation [31,67–70], spin pumping [4,60,67,71,72] or spin Seebeck effect [4,61,73,74]. See Ref. [26] for additional discussion.

*Experimental fits and discussion.*–We now use the above equations to derive the surface magnetic properties of the LCO film. Our transport measurements suggest that the magnetic moments of the LCO/Pt interface have a PM-like response. Given the observed field-dependence of the magnetoresistance in LCO/Pt, we assume that the Pt electrons interact with the spins of a low-dimensional Heisenberg FM –a system whose magnetic response is similar to a PM with $T_c$=0 and a large effective spin due to short-range FM interactions [75]– and compute the spin-correlations $\langle S^2_{\parallel,\perp}\rangle$ and $\langle S_\parallel \rangle$ that enter Eqs. (2) using the well-established random phase approximation [76]. For the fitting of the experimental data we considered that the Co atoms at the surface can exhibit either two-dimensional (2D) or one-dimensional (1D) FM exchange coupling $J$, that $S$ can be any of the possible ones in the $d$-shell (except 0 and ½, which result in no magnetoresistance correction), considered different spin coverage $\eta$ ($n_s$=/$\eta$ $a^2_{LCO}$ with $a_{LCO}$=3.904Å the LCO lattice constant [46]), and assumed collinear $s$-$d$ exchange coupling given that $\rho(H_x) \approx \rho(H_z)$ (Fig. 2b) [26]. $D$ was calculated using Einstein's relation $D$=1/2$e^2\rho\nu$, and $\theta_{SH}$ and $\lambda$ of Pt were estimated from the measured $\rho(T)$ [26,77]. Excellent fits to the FDMR measurements [Figs. 2(a)-2(b)] were found for a large range of parameters, some of which are summarized in Tables S1-S2 [26]. Figures 3(a)-3(b) show representative fits, evidencing the extraordinary good agreement with the experiment.

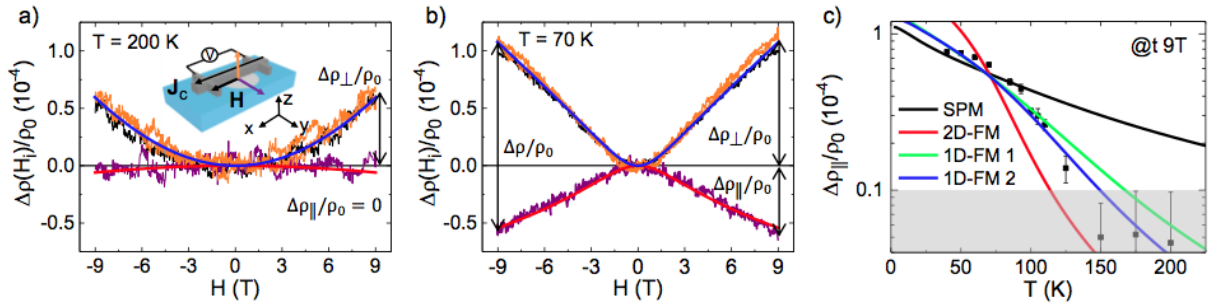

FIG. 3. (a)-(b) Red and blue lines are representative fits of the experimental data shown in Figs. 2(a)-2(b) calculated using Eqs. (2) and (3). Here we model the LCO surface as a plane of 1D-FM Co chains with $S$=3/2, $\eta$=1/3 and $J$=13.1meV, and used $\nu J_{sd}$=–0.15 (constant with T), $\theta_{SH}$=0.115(0.098), $\lambda$=4.00nm(4.72nm) and $D$=70.9·$10^{-6}$m$^2$s$^{-1}$(83.7·$10^{-6}$m$^2$s$^{-1}$) at 200K(70K). We assume $g$=2. (c) Fits of the experimental data $\Delta\rho_\parallel(T)/\rho_0$ [Fig. 2(c)] obtained modeling the LCO/Pt interface as a 2D and a 1D

Heisenberg FM (2D-FM and 1D-FM 1, respectively), as 1D-FM ladders (1D-FM 2), and as a SPM. See Tables S1-S3 [26] for details of the fitting parameters.

The temperature dependence of $\frac{\Delta\rho_\parallel}{\rho_0}(T)$ provides additional information about the magnetic ordering of the LCO surface through $G_s(\langle S_\perp^2 \rangle)$. Fig. 3(c) shows the best fits obtained for the experimental data $\frac{\Delta\rho_\parallel}{\rho_0}(T)$ modeling the LCO surface as a 2D Heisenberg magnet, a plane of spin chains, and a plane of interacting spin chains (spin ladders), all with FM coupling between spins. We also consider the case of a superparamagnet (SPM), which is described by exhibiting zero Heisenberg exchange coupling and large effective *S*. Our analysis evidences that a SPM cannot describe the temperature dependence of the magnetoresistance in LCO/Pt and confirms our assumption that the surface of LCO behave as a low-dimensional Heisenberg FM [see Fig. 3(c)]. Based on the origin of the FM coupling in LCO, we believe that the most plausible surface magnetic state in LCO films is a 2D Heisenberg FM originated by a HS-LS-HS $Co^{2+}$-$Co^{3+}$-$Co^{2+}$ superexchange interaction, provided that the surface arrangement is of the checkboard type (see Ref. [26] for a detailed discussion). The formation of stripe domains at the surface of tensile-strained LCO films has been observed experimentally [45,78] and confirmed in numerical calculations [78,79], making thus plausible that the LCO surface exhibits 1D FM ordering as well. Although we cannot unambiguously distinguish between the 1D and 2D FM cases from the magnetoresistance measurements, our analysis suggests that the surface Co ions might exhibit both contributions because the temperature dependence is fitted best by the model of spin ladders [blue line, Fig. 3(c)].

*Conclusions.*–Our SMR measurements in STO/LCO/Pt structures, together with the presented theoretical model, provide a clear evidence that the surface of LCO thin films is magnetically decoupled from its bulk and behaves as a low-dimensional FM system. Our microscopic model of the magnetoresistance in MI/NM bilayers revises the current SMR theory by introducing $G_s$, integrates both SMR and HMR contributions in the same set of analytical equations, and provides a simple way to correlate the magnetic properties of the MI through the spin conductances $G_{r,i,s}$ while covering a wide range of the magnetic order. Our theory sets the base for a better understanding of diverse spin transport phenomena involving MI/NM interfaces, and their manifestation on transport properties, as well as can help to address questions related to quantum magnetism or skyrmions.

*Acknowledgments.*–We thank Emilio Artacho for fruitful discussions. This work was supported by the European Union 7th Framework Programme under the Marie Curie Actions (607904-13-SPINOGRAPH), the European Research Council (257654-SPINTROS), and the European Regional Development Fund (ERDF), by the Spanish Ministry of Economy, Industry and Competitiveness (MAT2015-65159-R, FIS2014-55987-P, MAT2013-46593-C6-4-P, MAT2016-80762-R and FIS2017-82804-P), by the Basque Government (UPV/EHU Project IT-756-13 and IT-621-13), by the Regional Council of Gipuzkoa (100/16), and by the Xunta de Galicia (Centro singular de investigación de Galicia accreditation 2016–2019, ED431G/09).

J.M.G.-P. thanks the Spanish MINECO for a Ph.D. fellowship (Grant No. BES-2016-077301).

**REFERENCES**

[1]   S. Maekawa and T. Shinjo, *Spin Dependent Transport in Magnetic Nanostructures* (Taylor and Francis, New York, 2002).

[2]   A. Kobs, S. Heße, W. Kreuzpaintner, G. Winkler, D. Lott, P. Weinberger, A. Schreyer, and H. P. Oepen, Phys. Rev. Lett. **106**, 217207 (2011).

[3]   A. Kobs and H. P. Oepen, Phys. Rev. B **93**, 014426 (2016).

[4]   M. Weiler, M. Althammer, M. Schreier, J. Lotze, M. Pernpeintner, S. Meyer, H. Huebl, R. Gross, A. Kamra, J. Xiao, Y.-T. Chen, H. J. Jiao, G. E. W. Bauer, and S. T. B. Goennenwein, Phys. Rev. Lett. **111**, 176601 (2013).

[5]   H. Nakayama, M. Althammer, Y.-T. Chen, K. Uchida, Y. Kajiwara, D. Kikuchi, T. Ohtani, S. Geprägs, M. Opel, S. Takahashi, R. Gross, G. E. W. Bauer, S. T. B. Goennenwein, and E. Saitoh, Phys. Rev. Lett. **110**, 206601 (2013).

[6]   C. Hahn, G. de Loubens, O. Klein, M. Viret, V. V. Naletov, and J. Ben Youssef, Phys. Rev. B **87**, 174417 (2013).

[7]   Y.-T. Chen, S. Takahashi, H. Nakayama, M. Althammer, S. T. B. Goennenwein, E. Saitoh, and G. E. W. Bauer, Phys. Rev. B **87**, 144411 (2013).

[8]   N. Vlietstra, J. Shan, V. Castel, B. J. van Wees, and J. Ben Youssef, Phys. Rev. B **87**, 184421 (2013).

[9]   N. Vlietstra, J. Shan, V. Castel, J. Ben Youssef, G. E. W. Bauer, and B. J. van Wees, Appl. Phys. Lett. **103**, 032401 (2013).

[10]  M. Althammer, S. Meyer, H. Nakayama, M. Schreier, S. Altmannshofer, M. Weiler, H. Huebl, S. Geprägs, M. Opel, R. Gross, D. Meier, C. Klewe, T. Kuschel, J.-M. Schmalhorst, G. Reiss, L. Shen, A. Gupta, Y.-T. Chen, G. E. W. Bauer, E. Saitoh, and S. T. B. Goennenwein, Phys. Rev. B **87**, 224401 (2013).

[11]  M. Isasa, A. Bedoya-Pinto, S. Vélez, F. Golmar, F. Sanchez, L. E. Hueso, J. Fontcuberta, and F. Casanova, Appl. Phys. Lett. **105**, 142402 (2014).

[12]  T. Kosub, S. Vélez, J. M. Gomez-Perez, L. E. Hueso, J. Faßbender, F. Casanova, and D. Makarov, Appl. Phys. Lett. **113**, 222409 (2018).

[13]  C. O. Avci, K. Garello, A. Ghosh, M. Gabureac, S. F. Alvarado, and P. Gambardella, Nat. Phys. **11**, 570 (2015).

[14]  C. O. Avci, K. Garello, J. Mendil, A. Ghosh, N. Blasakis, M. Gabureac, M. Trassin, M. Fiebig, and P. Gambardella, Appl. Phys. Lett. **107**, 192405 (2015).

[15]  C. O. Avci, J. Mendil, G. S. D. Beach, and P. Gambardella, Phys. Rev. Lett. **121**, 087207 (2018).

[16]  H. Nakayama, Y. Kanno, H. An, T. Tashiro, S. Haku, A. Nomura, and K. Ando, Phys. Rev. Lett. **117**, 116602 (2016).

[17]  J. Kim, P. Sheng, S. Takahashi, S. Mitani, and M. Hayashi, Phys. Rev. Lett. **116**, 097201 (2016).

[18] M. Althammer, J. Phys. D. Appl. Phys. **51**, 313001 (2018).

[19] M. I. Dyakonov and V. I. Perel, Phys. Lett. A **35**, 459 (1971).

[20] J. E. Hirsch, Phys. Rev. Lett. **83**, 1834 (1999).

[21] A. Hoffmann, IEEE Trans. Magn. **49**, 5172 (2013).

[22] J. Sinova, S. O. Valenzuela, J. Wunderlich, C. H. Back, and T. Jungwirth, Rev. Mod. Phys. **87**, 1213 (2015).

[23] V. M. Edelstein, Solid State Commun. **73**, 233 (1990).

[24] A. Manchon, H. C. Koo, J. Nitta, S. M. Frolov, and R. A. Duine, Nat. Mater. **14**, 871 (2015).

[25] X. Jia, K. Liu, K. Xia, and G. E. W. Bauer, Europhys. Lett. **96**, 17005 (2011).

[26] See Supplemental Material at [URL to be inserted by publisher] for additional experimental data, further discussions regarding the expected Co surface magnetic states, surface magnetic arrangement and magnetic decoupling of the LCO surface, the formal derivation of Eqs. (1) and (2), and complementary information and discussion regarding the modeling and fitting parameters.

[27] M. Isasa, S. Vélez, E. Sagasta, A. Bedoya-Pinto, N. Dix, F. Sánchez, L. E. Hueso, J. Fontcuberta, and F. Casanova, Phys. Rev. Appl. **6**, 034007 (2016).

[28] A. Aqeel, N. Vlietstra, J. A. Heuver, G. E. W. Bauer, B. Noheda, B. J. van Wees, and T. T. M. Palstra, Phys. Rev. B **92**, 224410 (2015).

[29] A. Aqeel, N. Vlietstra, A. Roy, M. Mostovoy, B. J. van Wees, and T. T. M. Palstra, Phys. Rev. B **94**, 134418 (2016).

[30] K. Ganzhorn, J. Barker, R. Schlitz, B. A. Piot, K. Ollefs, F. Guillou, F. Wilhelm, A. Rogalev, M. Opel, M. Althammer, S. Geprägs, H. Huebl, R. Gross, G. E. W. Bauer, and S. T. B. Goennenwein, Phys. Rev. B **94**, 094401 (2016).

[31] S. Vélez, A. B. Pinto, W. Yan, L. E. Hues, and F. Casanova, Phys. Rev. B **94**, 174405 (2016).

[32] D. Hou, Z. Qiu, J. Barker, K. Sato, K. Yamamoto, S. Vélez, J. M. Gomez-Perez, L. E. Hueso, F. Casanova, and E. Saitoh, Phys. Rev. Lett. **118**, 147202 (2017).

[33] J. M. Gomez-Perez, S. Vélez, L. McKenzie-Sell, M. Amado, J. Herrero-Martín, J. López-López, S. Blanco-Canosa, L. E. Hueso, A. Chuvilin, J. W. A. Robinson, and F. Casanova, Phys. Rev. Appl. **10**, 044046 (2018).

[34] V. G. Bhide, D. S. Rajoria, G. Rama Rao, and C. N. R. Rao, Phys. Rev. B **6**, 1021 (1972).

[35] M. A. Señarís-Rodríguez and J. B. Goodenough, J. Solid State Chem. **116**, 224 (1995).

[36] P. Ravindran, P. A. Korzhavyi, H. Fjellvag, and A. Kjekshus, Phys. Rev. B **60**, 16423 (1999).

[37] J. Androulakis, N. Katsarakis, and J. Giapintzakis, Phys. Rev. B **64**, 174401 (2001).

[38] J.-Q. Yan, J.-S. Zhou, and J. B. Goodenough, Phys. Rev. B **70**, 014402 (2004).

[39] A. M. Durand, D. P. Belanger, C. H. Booth, F. Ye, S. Chi, J. A. Fernandez-Baca, and M. Bhat, J. Phys. Condens. Matter **25**, 382203 (2013).

[40] A. Ikeda, T. Nomura, Y. H. Matsuda, A. Matsuo, K. Kindo, and K. Sato, Phys. Rev. B **93**, 220401 (2016).

[41] J. Buckeridge, F. H. Taylor, and C. R. A. Catlow, Phys. Rev. B **93**, 155123 (2016).

[42] H. Hsu, P. Blaha, and R. M. Wentzcovitch, Phys. Rev. B **85**, 140404 (2012).

[43] D. Fuchs, E. Arac, C. Pinta, S. Schuppler, R. Schneider, and H. v. Löhneysen, Phys. Rev. B **77**, 014434 (2008).

[44] S. Park, R. Ryan, E. Karapetrova, J. W. Kim, J. X. Ma, J. Shi, J. W. Freeland, and W. Wu, Appl. Phys. Lett. **95**, 072508 (2009).

[45] W. S. Choi, J. Kwon, H. Jeen, J. E. Hamann-Borrero, A. Radi, S. Macke, R. Sutarto, F. He, G. A. Sawatzky, V. Hinkov, M. Kim, and H. N. Lee, Nano Lett. **12**, 4966 (2012).

[46] F. Rivadulla, Z. Bi, E. Bauer, B. Rivas-Murias, J. M. Vila-Fungueiriño, and Q. Jia, Chem. Mater. **25**, 55 (2013).

[47] L. Qiao, J. H. Jang, D. J. Singh, Z. Gai, H. Xiao, A. Mehta, R. K. Vasudevan, A. Tselev, Z. Feng, H. Zhou, S. Li, W. Prellier, X. Zu, Z. Liu, A. Borisevich, A. P. Baddorf, and M. D. Biegalski, Nano Lett. **15**, 4677 (2015).

[48] B. Rivas-Murias, I. Lucas, P. Jiménez-Cavero, C. Magén, L. Morellón, and F. Rivadulla, Nano Lett. **16**, 1736 (2016).

[49] Y. Yokoyama, Y. Yamasaki, M. Taguchi, Y. Hirata, K. Takubo, J. Miyawaki, Y. Harada, D. Asakura, J. Fujioka, M. Nakamura, H. Daimon, M. Kawasaki, Y. Tokura, and H. Wadati, Phys. Rev. Lett. **120**, 206402 (2018).

[50] Y.-T. Chen, S. Takahashi, H. Nakayama, M. Althammer, S. T. B. Goennenwein, E. Saitoh, and G. E. W. Bauer, J. Phys. Cond. Matter. **28**, 103004 (2016).

[51] M. I. Dyakonov, Phys. Rev. Lett. **99**, 126601 (2007).

[52] S. Vélez, V. N. Golovach, A. Bedoya-Pinto, M. Isasa, E. Sagasta, M. Abadia, C. Rogero, L. E. Hueso, F. S. Bergeret, and F. Casanova, Phys. Rev. Lett. **116**, 016603 (2016).

[53] A. Brataas, Y. V. Nazarov, and G. E. W. Bauer, Phys. Rev. Lett. **84**, 2481 (2000).

[54] J. Flipse, F. K. Dejene, D. Wagenaar, G. E. W. Bauer, J. Ben Youssef, and B. J. van Wees, Phys. Rev. Lett. **113**, 027601 (2014).

[55] J. Xiao and G. E. W. Bauer, ArXiv:1508.02486 (2015).

[56] T. Shang, Q. F. Zhan, H. L. Yang, Z. H. Zuo, Y. L. Xie, Y. Zhang, L. P. Liu, B. M. Wang, Y. H. Wu, S. Zhang, and R.-W. Li, Phys. Rev. B **92**, 165114 (2015).

[57] J. M. Pruneda, V. Ferrari, R. Rurali, P. B. Littlewood, N. A. Spaldin, and E. Artacho, Phys. Rev. Lett. **99**, 226101 (2007).

[58] S. Venkatraman and A. Manthiram, Chem. Mater. **14**, 3907 (2002).

[59] H. A. Tahini, X. Tan, U. Schwingenschlögl, and S. C. Smith, ACS Catal. **6**, 5565 (2016).

[60] Y. Shiomi and E. Saitoh, Phys. Rev. Lett. **113**, 266602 (2014).

[61] S. M. Wu, J. E. Pearson, and A. Bhattacharya, Phys. Rev. Lett. **114**, 186602 (2015).

[62] S. Meyer, M. Althammer, S. Geprägs, and S. T. B. Goennenwein, Appl. Phys. Lett. **104**, 242411 (2014).

[63] S. R. Marmion, M. Ali, M. McLaren, D. A. Williams, and B. J. Hickey, Phys. Rev. B **89**, 220404 (2014).

[64] K. Uchida, Z. Qiu, T. Kikkawa, R. Iguchi, and E. Saitoh, Appl. Phys. Lett. **106**, 052405 (2015).

[65] R. Schlitz, T. Kosub, A. Thomas, S. Fabretti, K. Nielsch, D. Makarov, and S. T. B. Goennenwein, Appl. Phys. Lett. **112**, 132401 (2018)

[66] T. Lino, T. Moriyama, H. Iwaki, H. Aono, Y. Shiratsuchi, and T. Ono, Appl. Phys. Lett. **114**, 022402 (2019).

[67] Y. Kajiwara, K. Harii, S. Takahashi, J. Ohe, K. Uchida, M. Mizuguchi, H. Umezawa, H. Kawai, K. Ando, K. Takanashi, S. Maekawa, and E. Saitoh, Nature **464**, 262 (2010).

[68] L. J. Cornelissen, J. Liu, R. A. Duine, J. Ben Youssef, and B. J. van Wees, Nat. Phys. **11**, 1022 (2015).

[69] S. T. B. Goennenwein, R. Schlitz, M. Pernpeintner, K. Ganzhorn, M. Althammer, R. Gross, and H. Huebl, Appl. Phys. Lett. **107**, 172405 (2015).

[70] R. Lebrun, A. Ross, S. A. Bender, A. Qaiumzadeh, L. Baldrati, J. Cramer, A. Brataas, R. A. Duine, and M. Kläui, Nature **561**, 222 (2018).

[71] Y. Tserkovnyak, A. Brataas, G. E. W. Bauer, and B. I. Halperin, Rev. Mod. Phys. **77**, 1375 (2005).

[72] E. Saitoh, M. Ueda, H. Miyajima, and G. Tatara, Appl. Phys. Lett. **88**, 182509 (2006).

[73] K. Uchida, J. Xiao, H. Adachi, J. Ohe, S. Takahashi, J. Ieda, T. Ota, Y. Kajiwara, H. Umezawa, H. Kawai, G. E. W. Bauer, S. Maekawa, and E. Saitoh, Nat. Mater. **9**, 894 (2010).

[74] S. Geprägs, A. Kehlberger, F. Della Coletta, Z. Qiu, E.-J. Guo, T. Schulz, C. Mix, S. Meyer, A. Kamra, M. Althammer, H. Huebl, G. Jakob, Y. Ohnuma, H. Adachi, J. Barker, S. Maekawa, G. E. W. Bauer, E. Saitoh, R. Gross, S. T. B. Goennenwein, and M. Kläui, Nat. Commun. **7**, 10452 (2016).

[75] N. D. Mermin and H. Wagner, Phys. Rev. Lett. **17**, 1133 (1966).

[76] N. Majlis, *The Quantum Theory of Magnetism*, 2nd ed. (World Scientific Publishing Co. Pte. Ltd, Singapore, 2007).

[77] E. Sagasta, Y. Omori, M. Isasa, M. Gradhand, L. E. Hueso, Y. Niimi, Y. Otani, and F. Casanova, Phys. Rev. B **94**, 060412 (2016).

[78] J.-H. Kwon, W. S. Choi, Y.-K. Kwon, R. Jung, J.-M. Zuo, H. N. Lee, and M. Kim, Chem. Mater. **26**, 2496 (2014).

[79] A. O. Fumega and V. Pardo, Phys. Rev. Mater. **1**, 054403 (2017).

# Spin Hall magnetoresistance in a low-dimensional magnetic insulator

## SUPPLEMENTAL MATERIAL

Saül Vélez[1,2,*], Vitaly N. Golovach[3,4,5,**], Juan M. Gomez-Perez[1], Cong Tinh Bui[6,7], F. Rivadulla[6], Luis E. Hueso[1,5], F. Sebastian Bergeret[3,4,†], and Fèlix Casanova[1,5,††]

[1]*CIC nanoGUNE, 20018 Donostia-San Sebastian, Basque Country, Spain*
[2]*Department of Materials, ETH Zürich, 8093 Zürich, Switzerland*
[3]*Centro de Física de Materiales (CFM-MPC), Centro Mixto CSIC-UPV/EHU, 20018 Donostia-San Sebastian, Basque Country, Spain*
[4]*Donostia International Physics Center (DIPC), 20018 Donostia-San Sebastian, Basque Country, Spain*
[5]*IKERBASQUE, Basque Foundation for Science, 48013 Bilbao, Basque Country, Spain*
[6]*Centro Singular de Investigación en Química Biolóxica e Materiales Moleculares (CiQUS), and Departamento de Química-Física, Universidad de Santiago de Compostela, 15782 Santiago de Compostela, Spain*
[7]*Department of Electrical and Electronic Engineering, Tokyo Institute of Technology, 2-12-1 Ookayama, Meguro, Tokyo 152-0033, Japan*

[*] saul.velez@mat.ethz.ch
[**] vitaly.golovach@ehu.eus
[†] sebastian_bergeret@ehu.eus
[††] f.casanova@nanogune.eu

## S1. Temperature dependence of the magnetic moment in the LaCoO$_3$ films

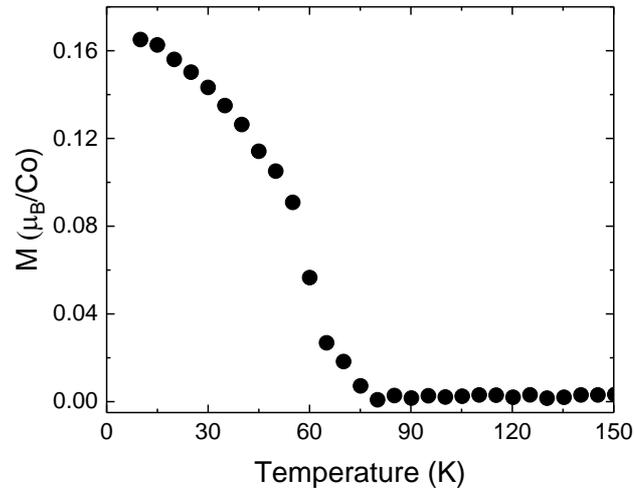

FIG. S1. Temperature dependence of the in-plane magnetic moment in the LaCoO$_3$(19nm) film in the presence of an in-plane magnetic field of 100 Oe.

## S2. Resistivity of the Pt layer

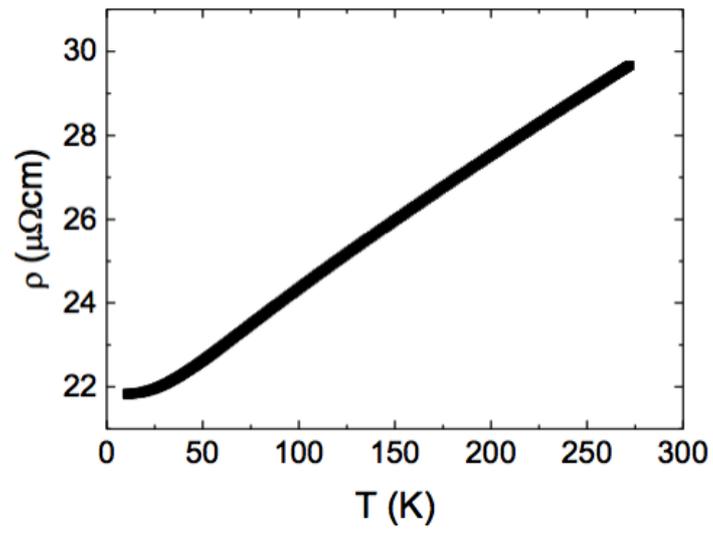

FIG. S2. Temperature dependence of the resistivity of the Pt(7nm) layer in LaCoO$_3$(19nm)/Pt(7nm).

## S3. Anomalous Hall-like measurements in LaCoO$_3$(19nm)/Pt(7nm)

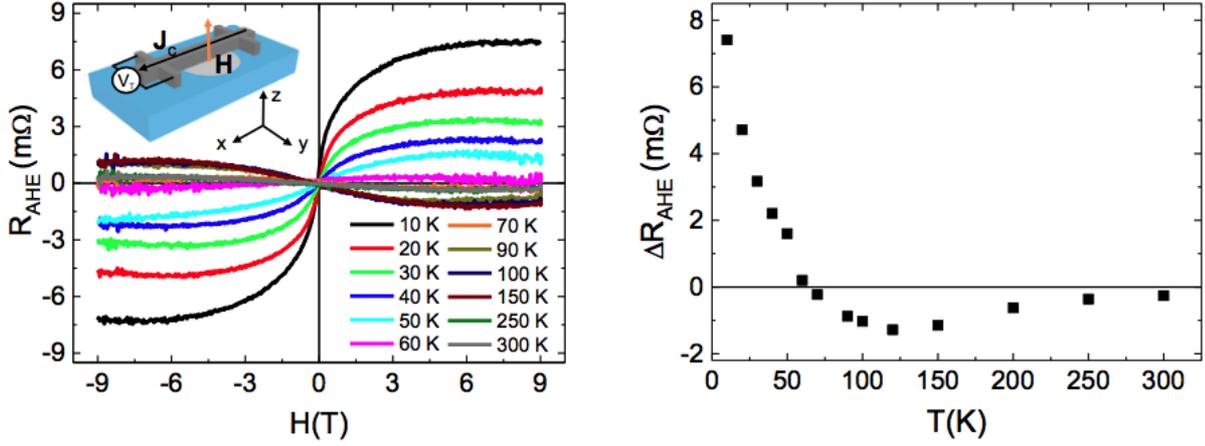

FIG. S3. (a) Magnetic field dependence of the anomalous Hall-like resistance $R_{AHE}(H_z)$ measured in LaCoO$_3$(19nm)/Pt(7nm) at different temperatures. $R_{AHE}$ is estimated from the Hall measurements ($V_T/I$, see sketch for the measurement configuration) after subtraction of the linear background associated to the ordinary Hall effect, following the same procedure as in Ref. [S1]. However, note that the surface of the LaCoO$_3$ thin film is not saturated at this magnetic field regime as we demonstrate in the main text [see for instance Figs. 1(g) and 2(b)]. Therefore, this procedure only allows us to estimate the anomalous Hall-like effect in our sample, and its temperature-dependence, for comparison with the data reported in Ref. [S1]. (b) Amplitude of the anomalous Hall resistance $\Delta R_{AHE}$ as a function of temperature. $\Delta R_{AHE}$ is calculated as $[R_{AHE}(9T)-R_{AHE}(-9T)]/2$ from the data in panel (a). The temperature dependence resembles the one reported in Ref. [S1].

## S4. Spin Hall Magnetoresistance as a probe for surface magnetization: strength of the technique

When dealing with the surface of magnetic insulators (MIs), either XAS or XPS measurements are not suitable to resolve the magnetic response, or oxidation state(s), of the very top surface magnetic atoms from the bulk properties, even when carried out in thin films. That is because of the relatively long penetration length of these techniques, where the bulk response saturates the signal, thus hindering the detection of the distinct properties the first atomic layer may exhibit. For instance, in the case of $LaCoO_3$/Pt, the surface exhibits a smooth magnetic behavior, which is difficult to discern on top of a significantly stronger and more abrupt magnetic response of the "bulk" of the film (which, in particular, shows hysteresis). For the same reason, standard magnetic surface techniques such as magneto-optical Kerr effect, magnetic force microscopy, or x-ray magnetic circular dichroism are not an option either. Other surface-sensitive techniques such as spin-polarized scanning tunneling microscopy or scanning electron microscopy with polarization analysis cannot be used in insulating substrates, thus becoming unpractical for measuring the surface magnetism in $LaCoO_3$. Only extremely complex, depth sensitive techniques, such as polarized neutron reflectometry might resolve the magnetic response of the surface of insulating films. However, such measurements require the use of large-scale facilities equipped with a neutron source. The strength of transport measurements in MI/NM bilayers, as the ones we performed in $LaCoO_3$/Pt, is that they allow for a relatively easy and fast direct access to the very surface magnetic properties of MIs [S2-S4].

## S5. Discussion regarding oxidation states of the surface Co atoms, magnetic ordering and decoupling of the surface in LaCoO$_3$ films

*Decoupling of the surface and expected oxidation and spin states of the surface Co atoms.*–The magnetic decoupling of the surface Co atoms from the rest of the thin film points to a weakening of the chemical bond between the Co atom of the surface CoO$_2$ plane and the oxide layers underneath. This might happen naturally due to a different coordination of the Co$^{3+}$ atom at the surface than in the bulk of the thin film. The polar nature of the LaCoO$_3$ crystal makes this material susceptible to ionic surface reconstructions [S6-S12], which can be accomplished by a change of the ionic state of a portion of Co atoms at the surface.

Since no DFT analysis of the LaCoO$_3$ surface has been performed, we can only make a guess about plausible surface Co states and interactions (that are compatible with the results obtained) based on the common sense and known literature. We went through a number of oxidation and spin states for the surface Co atoms before settling with what we think is the most plausible scenario.

First, transformation of part of Co$^{3+}$ atoms to Co$^{2+/4+}$, plus the formation of oxygen vacancies, is a likely scenario to occur at the surface of LaCoO$_3$. The reduction of coordination and the presence of oxygen vacancies inherent to the surface of LaCoO$_3$ are expected to lead to a change of the oxidation and spin state of the Co atoms in a similar way as it does for the spin state of Mn$^{3+/4+}$ in La$_{0.7}$Sr$_{0.3}$MnO$_3$ [S13,S14]. Support for this large hybridization is given by the immediate formation of oxygen vacancies in Li$_x$CoO$_2$, La$_{1-x}$Sr$_x$CoO$_3$, etc, after Co$^{3+}$ to Co$^{4+}$ transformation [S15]. Moreover, the relevance of Co$^{2+}$ in LaCoO$_3$ has been demonstrated in [S16]. On the other hand, surface adsorption of an oxygen atom (from moisture and/or molecular oxygen) would oxidize surface Co to Co$^{4+}$. Now, if we assume that low surface coordination will favor higher-spin states, we end up with a number of possible combinations of oxidation and spin states for Co$^{2+}$ (3/2), Co$^{3+}$ (1, 2), and Co$^{4+}$ (3/2, 5/2). Note that, according to our theory, a MR is only expected for S=1 or larger. Therefore, our transport measurements directly reveal that contribution from high-spin (HS) states at the surface must be large. But it is also worth noting that we are not necessarily restricted only to the presence of HS states.

Surface cobalt ions having intermediate spin (IS) Co$^{3+}$ could show ferromagnetism induced by a vibronic e1–O–e0 superexchange. However, the SMR is absent in the case of spin 1/2. Therefore, a transition to a higher spin state would be required to account for the MR. In this case, antiferromagnetic ordering between surface HS Co$^{3+}$ could give a weak canted-spin ferromagnetic moment. However, the large surface magnetization observed in single crystals of LaCoO$_3$ in Ref. [S17], suggested this possibility is less likely (see also comments below regarding the modeling and a possible anisotropic exchange interaction and non-collinear magnetic ordering at the surface of LaCoO$_3$ due to the presence of Pt).

Finally, the presence of Co$^{2+}$ (S=3/2) ions associated to oxygen vacancies was corroborated in LaCoO$_3$ [S16]. In this case, a ferromagnetic interaction between 3/2-spins can occur via the spin-less Co$^{3+}$ nodes (low spin, LS, state), provided the arrangement of the two Co species on the surface is of the checkerboard type. Given the different sizes of these ions, the reduction of the elastic energy associated to a cooperative ionic ordering could indeed favor such an ionic arrangement. Therefore, our transport measurements indicate that the most plausible and dominant magnetic state of the surface of LaCoO$_3$ is a low-dimensional Heisenberg FM mediated by HS-LS-HS Co$^{2+}$-Co$^{3+}$-Co$^{2+}$ interaction (see discussion below for more details).

*Expected surface magnetic reconstruction.*–Before discussing the spin states expected for the surface Co atoms, we need to agree on a scenario for the surface reconstruction. We pick the Co$^{2+}$-Co$^{3+}$ scenario as discussed above (a similar conclusion is reached for the Co$^{4+}$-Co$^{3+}$ scenario). The crystal-field splittings for the Co$^{2+}$ atom in that scenario is of the square planar type, whereas for the Co$^{3+}$ atom is of the square pyramidal type, see Figures S4(a) and (b). If we had a surface consisting of a checkerboard arrangement of Co$^{2+}$ in the HS (S=3/2) state and Co$^{3+}$ in the IS (S=1) state, then the interactions between these two species would be antiferromagnetic and strong. Indeed, the $d_{z^2}$ and $d_{xy}$ orbitals on both Co$^{2+}$ and Co$^{3+}$ would be half filled, leading to a strong antiferromagnetic superexchange between Co$^{2+}$ and Co$^{3+}$. As a result, the LaCoO$_3$ surface would be a 2D ferrimagnet, consisting of two intercalated S=3/2 and S=1 lattices. In principle, we could consider such a model as well (although perhaps only in some limiting cases since the RPA theory for such a model is rather complicated), in addition to the models we have considered already. However, there is one contradictory circumstance: In order to fit the experimental data, we do not need such a strong superexchange coupling. Our fits use a spin-spin coupling which is compatible by order of magnitude with the one present in the bulk of the thin film (i.e. which gives a $T_c$ of about 85 K). We believe that, both in the bulk of the thin film and on the surface, the superexchange is mediated by spin-less Co$^{3+}$ node, or at least by very similar superexchange processes. And since no ferrimagnetism is observed for the bulk of the thin film (only ferromagnetic behavior is reported by all groups), we tend to believe that the superexchange goes via spin-less Co$^{3+}$ nodes both in the bulk of the thin film and on its surface.

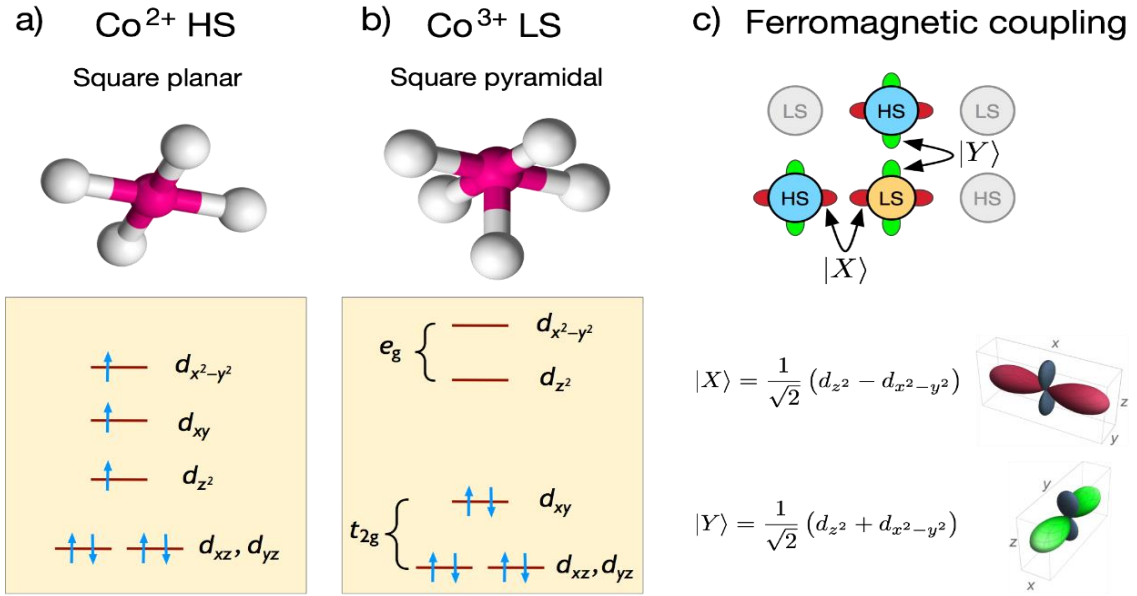

FIG. S4. A mechanism of ferromagnetic coupling at the surface of the LaCoO$_3$ film. As a result of the surface reconstruction, the Co atoms form an alternating pattern of Co$^{2+}$ (with an oxygen vacancy underneath) and Co$^{3+}$ in a 1:1 stoichiometry. (a) The crystal field at the Co$^{2+}$ atom has approximately a square planar symmetry. The d-shell multiplet splits moderately, because the smaller positive charge of the Co ion results in a larger Co-O distance. The spin configuration should thus be HS (S=3/2). (b) The crystal field at the Co$^{3+}$ atom has a square pyramidal symmetry. The d-shell multiplet splits stronger as compared to the previous case, due to the relatively shorter Co-O distance and hence a stronger covalent bonding which moves the $e_g$ and $t_{2g}$ multiplets further apart in energy from each other. The spin configuration should thus be LS (S=0). (c) Ferromagnetic coupling arises between nearest HS Co$^{2+}$ atoms due to the 90-degree-oriented |X> and |Y> orbitals of the $e_g$ multiplet. The LS Co$^{3+}$ atom mediates the ferromagnetic coupling, which occurs due to the exchange interaction of electrons (Hund's rule) on Co$^{3+}$ during the virtual processes of superexchange.

This hypothesis is reinforced by the mechanism of ferromagnetic coupling shown in Figure S4(c). The $e_g$ states ($d_{z^2}$ and $d_{x^2-y^2}$) can be combined during virtual transitions into two orthogonal combinations (X and Y), elongated either along the *x* and *y* directions, see Figure S4(c). The spin-less Co$^{3+}$ node allows, therefore, electrons from HS Co$^{2+}$ atoms to hop virtually on different X and Y orbitals and interact via Hund's rule. The oxygens (not shown in Figure S4(c)) mediate the hopping, which is somewhat stronger for the $e_g$ orbitals than for the $t_{2g}$ orbitals, because of the sigma-type bonding of the oxygens with the $e_g$ multiplet versus the π-type bonding with the $t_{2g}$ multiplet. Thus, under generic conditions, a ferromagnetic interaction is mediated between HS $Co^{2+}$ atoms which are oriented 90 degrees with respect to the spin-less $Co^{3+}$ node.

As for the $Co^{2+}$ atoms which are oriented 180 degrees, the situation is more complicated and the sign of the interaction is less obvious to tell because of competing interactions. On the one hand, the half-filled $e_g$ orbitals of $Co^{2+}$ favor antiferromagnetic superexchange via the spin-less $Co^{3+}$ node, which has empty $e_g$ orbitals. On the other hand, there is strong evidence in the literature (see, e.g. Ref. [S5] and a series of old papers by Goodenough cited therein) that the spin-less $Co^{3+}$ node favors a superexchange process during which first a $t_{2g}$ electron (in our case from $d_{xy}$) leaves, changing $Co^{3+}$ to $Co^{4+}$, and then a $e_g$ electron comes, restoring the ionic state back to $Co^{3+}$, but in a different spin state. This superexchange process is shown in Figure S5. The total interactions between 180 degree atoms can be dominated by this process, and thus be ferromagnetic, provided $Co^{3+}$ is close in energy to transit to $Co^{4+}$, i.e. the extraction energy of an electron from $Co^{3+}$ is sufficiently small to favor first extracting an electron and then adding the other during the virtual process. This is known to be the case for $Co^{3+}$ in the bulk of the thin film. At the surface, $Co^{3+}$ will have somewhat different electron extraction and addition energies, but since these energies enter in the denominator of the superexchange, we hope that the changes are not large enough to change the sign of the spin-spin interaction.

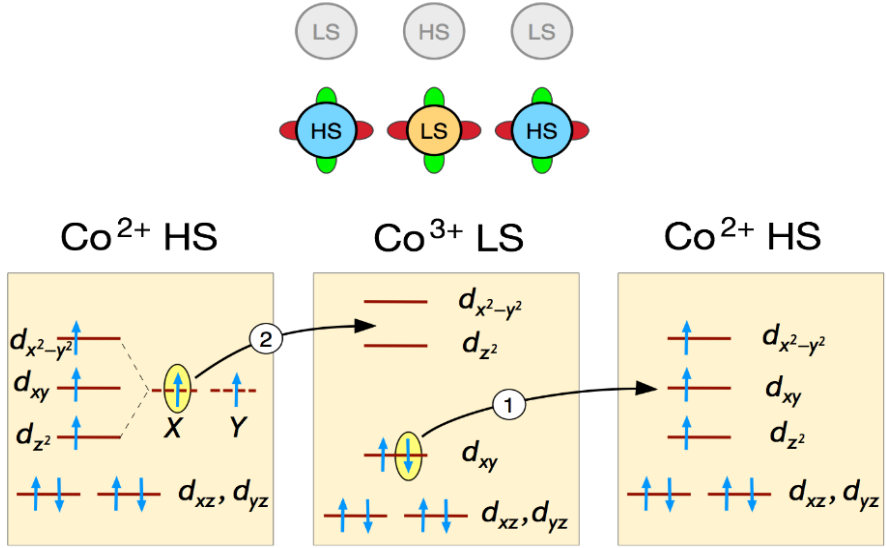

FIG. S5. Mechanism of ferromagnetic coupling between HS $Co^{2+}$ sites which are 180 degrees oriented. If the extraction energy of an electron from the LS $Co^{3+}$ site is sufficiently small, the superexchange is dominated by a virtual process which consists of, first, transferring a $t_{2g}$ electron from $Co^{2+}$ to $Co^{3+}$, and then, transferring an $e_g$ electron from a different $Co^{3+}$ to the same $Co^{2+}$. The Hund's rule on $Co^{3+}$ favors then a ferromagnetic interaction, which is expected to be sufficiently strong to compete with the antiferromagnetic Anderson superexchange. For the latter, it is required that either two $t_{2g}$ electrons leave $Co^{3+}$ or two $e_g$ electrons come on $Co^{3+}$, making the process energetically less favorable (i.e. far from resonances). This mechanism works also for 90-degrees-oriented HS-LS-HS sites, for which the antiferromagnetic Anderson superexchange does not occur and hence the ferromagnetic coupling is expected to be stronger for the 90-degrees-oriented than for the 180-degrees-oriented sites.

*Final remarks regarding the expected surface reconstruction.*–The reason why we assumed that every second Co atom is in the LS (S=0) state and mediates a ferromagnetic coupling between HS Co atoms goes back to the assumptions usually made in the literature on ferromagnetic coupling in $LaCoO_3$. In order for the Co atoms to interact ferromagnetically between themselves, the superexchange must occur between orbitals which are 90 degrees and not 180 degrees oriented with respect to each other (Goodenough-Kanamori rules). Such an assumption about the emergence of a checker-board pattern is made in the literature for the "bulk" of the thin film, see, e.g., Ref. [S5]. Actually, in all scenarios of ferromagnetism in LaCoO$_3$ thin films, some of the Co atoms are assumed to be in the LS (S=0) state. This is supported by experiments that show that the average magnetization per Co atom in the LaCoO$_3$ film is rather small (about 0.8 Bohr magneton per Co atom). In this context, we formulated the simplest possible scenario for ferromagnetic interactions at the surface, given the experimental evidence for the magnetic decoupling. We took into account the lowering of the crystal-field symmetry (from octahedral to square pyramidal) when working out the surface reconstruction scenario (LaCoO$_3$ is a polar material) and the stabilization of the LS state of Co$^{3+}$. Unfortunately, we cannot assert that this scenario is really occurring in the experiment, but that status have literally all scenarios put forward in attempts to explain the ferromagnetism of LaCoO$_3$ thin films. And that is despite the fact that there are many more experiments probing the magnetism of the thin film than the magnetism of the very surface. The SMR/HMR technique is unique in this sense, since its penetration depth is very small, basically given by the tail of the wave function of the Pt electron penetrating in the band gap of the LaCoO$_3$ insulator, and thus being only sensitive to very surface. However, the SMR/HMR technique cannot tell so precisely about the magnetic arrangement in the lateral direction on the surface, since the signal is collected from the surface as a whole and gets, consequently, averaged. Therefore, we are certain only about the magnetic decoupling of the surface, but we cannot discriminate so well between different scenarios of LS/HS alternation on the surface (such as checker-board versus stripes). This question remains to most part open in our work. We attempted to address it indirectly, via the temperature dependence of the SMR signal, considering a two-dimensional and parallel one-dimensional models. The differences between different curves are, unfortunately, not sufficiently large to make a firm statement about the distribution of HS and LS Co atoms on the surface. Hopefully, some other technique will be able to answer this question in the future

***Possible role of Pt at the LaCoO$_3$/Pt interface: spin-orbit coupling induced non-collinear magnetism and RKKY interaction.*** –As for the possibility of other magnetic states, such as AFM or non-collinear magnetic structures, we are confident that our experimental data is consistent with the collinear FM state. We can immediately discard the AFM ordering given that an AFM state would result in a somewhat opposite behavior for the SMR+HMR curves than the one measured, i.e. instead of being sharper than the paramagnetic case, they should then be flatter: From the magnetic field dependence in Fig. 2(b) of the main text, we see that the curves have a rather small characteristic scale (of about 2 T) for changing behavior from quadratic to roughly linear. The linear regime continues and we do not see saturation within the available data rage (H<9 T). This behavior is consistent with ferromagnetic correlations, which enhance the susceptibility to the magnetic field. Without ferromagnetic correlations we would expect the quadratic behavior to persist to about 50 T at T=70 K (i.e. $g\mu_B|H| = k_B T$, with $k_B$ the Boltzmann constant). In the AFM case, the quadratic behavior should persist to even larger values of the magnetic field.

Besides, the quasi-identical magnetic field response of $\rho_{xx}$ for the magnetic field applied along *x*- and along *z*-directions (black and orange curves in Fig. 2(b), respectively) indicate that isotropic exchange interaction must be dominating in our system, which is consistent with the expected *s-d* coupling between the itinerant electrons in Pt and the *d*-orbitals of the electrons in Co. Although we do not know the exact origin of the tiny MR difference observed (for $H||z$ with respect to $H||x$), we cannot completely exclude that magnetic anisotropy induced by the presence of Pt at the LaCoO$_3$/Pt interface might play a role here. For instance, the spin-orbit coupling of Pt might promote non-collinear magnetic structures in LaCoO$_3$/Pt in the range of small magnetic fields. However, in a field $H \sim 10$ T, these structures should probably be straightened out already. More research needs to be done in this direction in order to determine whether non-collinear structures are present in LaCoO$_3$/Pt, for instance, by employing local probes such as single nitrogen-vacancy defects in diamond to map the stray fields emanating from non-collinear structures [S18]. Our method provides the basis for further work in this direction.

Besides, the itinerant electrons in the Pt layer can also mediate a spin-spin coupling between the surface Co atoms in the LaCoO$_3$ film, giving rise to RKKY interaction. In a clean metal that interaction can be either ferromagnetic or antiferromagnetic, oscillating and decaying with the distance between local moments at the surface. In a strongly disordered metal, as it is Pt in our case, the RKKY interaction should have a non-vanishing average value only at distances smaller than or comparable with the Fermi wavelength in Pt. At such small distances the RKKY interaction is equivalent to Zener's double exchange and can be considered as an alternative to the superexchange mechanism discussed above. We estimate that for values of $\nu J_{sd} \sim 0.1$ and $\nu$ of about 1/eV per unit cell, the RKKY interaction ($J_{RKKY} \sim \nu J_{sd}^2$) is on the order of 10 meV. This value should further be reduced due to the suppression with the distance between local moments, but it is otherwise by order of magnitude in the range of the required values to fit the experimental data. We thus conclude that the RKKY mechanism may also be contributing to the observed ferromagnetic correlations at the surface.

## S6. Microscopic modeling of the magnetic insulator/normal metal interface and integrated Spin Hall and Hanle magnetoresistance corrections: physical meaning of $G_r$, $G_i$ and $G_s$, applicability of the model and modeling non-collinear magnetic exchange coupling in LaCoO$_3$/Pt

*Microscopic description of the SMR and integrated SMR and HMR transport equations.*–The scope of this work is to introduce the SMR/HMR technique and to relate the so far phenomenological parameters $G_s$, $G_r$, and $G_i$ to the magnetic state of the spins at the surface of the MI. In Section S8, we provide the full rigorous derivation of Eqs. (1) and (2) of the main text, which elucidates the physical origin of $G_s$ as well as of $G_r$ and $G_i$ in terms of the transverse and longitudinal spin relaxation rates and surface-induced exchange splitting [see Eq. (S98), Section S8]. Thus, we can see that $G_s$ is determined by the ability of the spin accumulation at the surface to relax by emitting a magnon into the MI. $G_r$ is determined by the anisotropy of spin relaxation in the metal at the surface with MI, whereas $G_i$ is determined by the exchange splitting induced by the MI onto the metal.

The derivation of Eq. (3) of the main text repeats one-to-one the derivation of the HMR equations, which we explained in great detail in the Supplemental Material of Ref. [S19]. The only difference here is that the boundary condition is more complicated (and less symmetric) than it was for HMR, leading to cumbersome intermediate expressions. Nevertheless, the end result in Eq. (3) of the main text is relatively compact, given that it combines SMR and HMR together and also generalizes SMR by including $G_s$. Notably, $G_r$ never enters alone in the end result, but appears always in combination with $G_s$ as $G_r$–$G_s$. One can see from Eq. (S98) of the Supplemental Material, Section S8, that the quantity $G_r$–$G_s$ is related to the transverse spin-relaxation rate. In the same way, the quantity $-G_s$ is related to the longitudinal spin-relaxation rate. Therefore, it is simpler to explain the physical meaning of $G_r$–$G_s$ than of $G_r$ alone. Only in the limit of strong dephasing $G_r$–$G_s$ can be approximated by $G_r$ and then $G_r$ appears alone in the end result as in the original SMR work. In most interesting cases, however, both spin relaxation and spin dephasing matter and our work gives a complete physical picture in terms of the longitudinal and transverse spin relaxation at the interface. This is why Eq. (3) of the main text does not only combine SMR and HMR together, but it also corrects SMR in an important way compared to the original work.

Besides, note that the SMR effect has been intensively studied experimentally in the recent years, but there was no practical way to relate the spin mixing conductances to the magnetic state of the interface. Although it was clear that the spin-mixing conductances can be derived microscopically from the scattering matrix of the electron for scattering off the interface, the derivation has never been carried out for quantum magnetic systems and the spin mixing conductances in MI/NM interfaces have been used as phenomenological parameters. We provide an alternative way to derive the spin-mixing conductances by relating them to the spin relaxation properties of the interface. This is an elegant and considerably more tractable way than calculating the elusive S-matrix of a many-body problem going the long way of relating it to the spin-mixing conductances in a disordered metallic system. In our SMR (+HMR) theory for MI/NM bilayers, we describe, for the first time, the spin mixing conductances from a microscopic point of view, as well as include the so-far-omitted effective spin conductance $G_S$, which have profound implications in both the charge and spin transport properties of the bilayer. For instance, our microscopic description of $G_{r,i,s}$ now allows addressing spin transport phenomena across magnetic phase transitions or to describe MR effects for different magnetic ordering of the MI layer, including paramagnets. More importantly, our theory can also be applied to describe any spin transport phenomena involving spin flow across interfaces, including spin pumping, spin Seebeck effect, or magnon spin transport, just to mention a few.

*Applicability of the model.*–In generic cases, one can compute the spin averages $\langle S^2_{\parallel,\perp} \rangle$ and $\langle S_{\parallel} \rangle$ by solving numerically the corresponding magnetic Hamiltonian or by employing approximate schemes, such as the random phase aproximation (RPA) [S20]. Equations (2) in the main text have been derived under the assumption that the mean free path for the conducting electrons between two consecutive scattering events at the localized moments is larger than the Fermi wavelength in the NM. Furthermore, we employed the so-called elastic approximation, assuming that the energy transfer between the local moments and itinerant electrons is negligible on the scale of the thermal smearing of the Fermi sea. The elastic approximation limits the applicability of the expressions for $G_{r,i,s}$ to sufficiently high temperatures, at which the self-consistent magnon band has a small width compared to the temperature. Therefore, these expressions are not applicable at temperatures well below $T_c$. Finally, we also dispensed with the effect of the conduction electrons on the magnetic configuration of the localized spins, assuming that the characteristic coupling energy between the latter is much larger than the exchange coupling $J_{sd}$.

***Modeling a non-collinear magnetic order at the LaCoO$_3$/Pt interface.***–As discussed in Section S5, our transport measurements indicate that collinear exchange coupling is dominating at the LaCoO$_3$/Pt interface. However, we cannot completely exclude that some contribution arising from anisotropic exchange coupling is present. In that situation, from the theory side, we could consider an exchange anisotropy interaction in the RPA model. In the case of easy-axis anisotropy, there is almost no modification to be made. It amounts to rewriting the magnon dispersion relation [now provided in Eq. (S13), Section S8] as $\hbar\omega_q = g\mu_B B - \langle S_\parallel\rangle(J_0^z - J_q)$, where $J_0^z$ is now different from $J_0$ due to the exchange anisotropy. Here, we denote by $J^z$ the coupling in front of $S^z S^z$ terms. The case of easy-axis anisotropy corresponds to making $J_0^z$ larger than it was in the SU(2)-symmetric case. The magnon spectrum acquires a gap already at $B=0$. This stabilizes the magnetization along $z$ and we would expect an enhancement of the SMR effect. Roughly, this corresponds to the slight anisotropy seen in the experiment. The orange curve ($H\|z$) in Fig. 2(b) tends to show a slightly larger SMR effect. But we are not sure that we would interpret the slight anisotropy correctly this way, because the effect is so small and we have not carried out a systematic study.

If the exchange anisotropy is of the easy-plane type, then the RPA theory can also be used, but the ground state needs to be chosen properly. Intuitively, one can understand that, if we make $J_z$ smaller than in the SU(2)-symmetric case, then the magnon frequency becomes negative for small $B$ and $q$. This unphysical result indicates that our choice of the order parameter is incorrect, which translates to an incorrect choice of the ground state. What may work, however, is to pick a generic order parameter on each local-moment site, i.e. the average value of the spin operator along a direction **n**. The unit vector **n** can even be different on each site. For a single site, the expressions for $S_+$ and $S_-$ are given incidentally below Eq. (S22) of Section S8. The RPA equation for the Green function can be written in the coordinate space in a straightforward manner, but to solve them in the Fourier space is more complicated and one has to make simplifying assumptions about the variation of the unit vector $n_j$ in space and about the average $<n_j S_j>$. Usually, it is sufficient to assume that $<n_j S_j>$ is the same on all sites, whereas $n_j$ changes smoothly. We have not attempted to implement the solution of the RPA equations for such a generic case, but it is clear physically that for a magnetic field in the plane of the film, the magnetization will be collinear with the field and the magnon spectrum will be given by $\hbar\omega_q = g\mu_B B - \langle S_\parallel\rangle[J_0 - (J_q^z + J_q)/2]$, which shows that the anisotropy stabilized the magnetization in the plane, because $J_0 > (J_q^z + J_q)/2$ in this case. Then, it is also clear that if the magnetic field is gradually oriented at an angle out of plane, the magnetization will lag behind and stop being collinear with the magnetic field. For the itinerant electrons this means that the Zeeman energy and the surface-induced exchange interaction are no longer collinear. Our SMR+HMR theory becomes invalid in this case and needs to be revised. This situation –where is considered an easy-axis and easy-plane anisotropy problem– is a quite complex one, going much beyond the scope of this work and could be treated separately in a follow up work.

## S7. Fitting of the experimental data

In the following we provide a set of fitting parameters of the experimental data considering that the Co atoms at the surface can exhibit either 2D and 1D FM exchange coupling $J$, that $S$ can be any of the possible ones in the $d$-shell (except 0 and ½, which result in no magnetoresistance correction), consider different spin coverage $\eta$ ($n_s = \eta/a^2_{LCO}$ with $a_{LCO}$=3.904Å the LaCoO$_3$ lattice constant [S21]), and allowed different sign and amplitude for the $J_{sd}$ coupling. Additionally, cases of interacting 1D-FM chains (spin ladders) and zero Heisenberg exchange coupling and large effective spin (SPM case) are also considered.

| $S$ | $T$ (K) | $\eta$ | $J$ (meV) | $\nu J_{sd}$ | $D$ ($10^{-6}$m$^2$s$^{-1}$) | $\theta_{SH}$ | $\lambda$ (nm) |
|---|---|---|---|---|---|---|---|
| 1 | 70 | 1/2 | 3.55 | 0.34 | 58.6 | 0.108 | 2.62 |
|   | 200 |     |      |      | 37.7 | 0.127 | 2.22 |
| 1 | 70 | 1 | 3.48 | 0.38 | 58.6 | 0.103 | 2.62 |
|   | 200 |   |      |      | 35.7 | 0.121 | 2.22 |
| 1 | 70 | 1/2 | 3.56 | -0.27 | 58.6 | 0.114 | 3.15 |
|   | 200 |     |      |       | 49.6 | 0.135 | 2.66 |
| 1 | 70 | 1 | 3.44 | -0.33 | 58.6 | 0.104 | 2.62 |
|   | 200 |   |      |       | 34.2 | 0.123 | 2.22 |
| 3/2 | 70 | 1/2 | 1.78 | 0.18 | 58.6 | 0.093 | 2.86 |
|     | 200 |     |      |      | 38.7 | 0.110 | 2.42 |
| 3/2 | 70 | 1 | 1.72 | 0.21 | 50.2 | 0.082 | 2.62 |
|     | 200 |   |      |      | 29.3 | 0.097 | 2.22 |
| 3/2 | 70 | 1/2 | 1.79 | -0.17 | 58.6 | 0.092 | 3.69 |
|     | 200 |     |      |       | 49.6 | 0.110 | 3.12 |
| **3/2** | **70** | **1** | **1.67** | **-0.13** | **58.6** | **0.102** | **2.82** |
|         | **200** |       |          |           | **37.2** | **0.121** | **2.39** |
| 2 | 70 | 1/2 | 1.06 | 0.13 | 58.6 | 0.084 | 3.16 |
|   | 200 |     |      |      | 42.2 | 0.100 | 2.68 |
| 2 | 70 | 1 | 1.02 | 0.15 | 46.3 | 0.075 | 2.62 |
|   | 200 |   |      |      | 27.0 | 0.088 | 2.23 |
| 2 | 70 | 1/2 | 1.06 | -0.12 | 58.6 | 0.084 | 3.96 |
|   | 200 |     |      |       | 49.6 | 0.100 | 3.35 |
| 2 | 70 | 1 | 1.01 | -0.10 | 46.3 | 0.085 | 2.95 |
|   | 200 |   |      |       | 32.0 | 0.100 | 2.49 |
| 5/2 | 70 | 1/2 | 0.69 | 0.11 | 55.0 | 0.079 | 3.23 |
|     | 200 |     |      |      | 40.9 | 0.094 | 2.73 |
| 5/2 | 70 | 1 | 0.66 | 0.11 | 44.0 | 0.071 | 2.62 |
|     | 200 |   |      |      | 26.0 | 0.083 | 2.22 |
| 5/2 | 70 | 1/2 | 0.69 | -0.10 | 58.6 | 0.080 | 4.12 |
|     | 200 |     |      |       | 49.6 | 0.095 | 3.49 |
| 5/2 | 70 | 1 | 0.66 | -0.09 | 40.9 | 0.076 | 3.01 |
|     | 200 |   |      |       | 30.8 | 0.090 | 2.55 |

TABLE S1. Parameters used to achieve a good agreement between theory and experiment considering that the surface of the LaCoO$_3$ film behaves as a 2D-FM. In bold are indicated the parameters used to compute the temperature dependence of $\Delta\rho_\parallel/\rho_0$ at 9T [see Fig. 3(c)] and the FDMR curves at 200 and 70K [see Figs. S3(a) and S3(b)] for the 2D-FM case.

| $S$ | $T$ (K) | $\eta$ | $J$ (meV) | $\nu J_{sd}$ | $D$ ($10^{-6}$m$^2$s$^{-1}$) | $\theta_{SH}$ | $\lambda$ (nm) |
|---|---|---|---|---|---|---|---|
| 1 | 70 | 1/3 | 35.0 | 0.34 | 50.2 | 0.105 | 2.62 |
| 1 | 200 | 1/3 | 35.0 | 0.34 | 42.5 | 0.124 | 2.22 |
| 1 | 70 | 1/3 | 30.6 | -0.25 | 87.9 | 0.118 | 4.06 |
| 1 | 200 | 1/3 | 30.6 | -0.25 | 74.4 | 0.140 | 3.43 |
| 3/2 | 70 | 1/3 | 13.8 | 0.16 | 70.3 | 0.098 | 3.40 |
| 3/2 | 200 | 1/3 | 13.8 | 0.16 | 59.5 | 0.116 | 2.88 |
| **3/2** | **70** | **1/3** | **13.1** | **-0.15** | **83.7** | **0.098** | **4.72** |
| **3/2** | **200** | **1/3** | **13.1** | **-0.15** | **70.9** | **0.115** | **4.00** |
| 2 | 70 | 1/3 | 7.21 | 0.11 | 70.3 | 0.097 | 3.40 |
| 2 | 200 | 1/3 | 7.21 | 0.11 | 59.5 | 0.114 | 2.88 |
| 2 | 70 | 1/3 | 6.77 | -0.11 | 79.9 | 0.089 | 5.04 |
| 2 | 200 | 1/3 | 6.77 | -0.11 | 67.6 | 0.105 | 4.26 |
| 5/2 | 70 | 1/3 | 4.26 | 0.080 | 70.3 | 0.097 | 3.40 |
| 5/2 | 200 | 1/3 | 4.26 | 0.080 | 59.5 | 0.114 | 2.88 |
| 5/2 | 70 | 1/3 | 3.91 | -0.086 | 79.9 | 0.085 | 5.26 |
| 5/2 | 200 | 1/3 | 3.91 | -0.086 | 67.6 | 0.101 | 4.45 |
| **3/2** | **70** | **2/3** | **3.97** | **-0.154** | **79.9** | **0.095** | **4.69** |
| **3/2** | **200** | **2/3** | **3.97** | **-0.154** | **67.6** | **0.112** | **3.97** |

TABLE S2. Rows 1 to 8: parameters used to achieve a good agreement between theory and experiment considering that the surface of the LaCoO$_3$ film behaves as 1D-FM Co chains. In bold are indicated the parameters used to compute the temperature dependence of $\Delta\rho_\parallel/\rho_0$ at 9T [see Fig. 3(c), green line] and the FDMR curves at 200 and 70K [see Figs. 3(a) and 3(b)] for the 1D-FM 1 case. Row 9, bolted: parameters used to achieve a good agreement between theory and experiment considering that the surface of the LaCoO$_3$ film behave as interacting 1D-FM chains (spin ladders). Case of 1D-FM 2 shown in Fig. 3(c), blue line.

| $S$ | $T$ (K) | $\eta$ | $J$ (meV) | $\nu J_{sd}$ | $D$ ($10^{-6}$m$^2$s$^{-1}$) | $\theta_{SH}$ | $\lambda$ (nm) |
|---|---|---|---|---|---|---|---|
| 29 | 70 | 1/2 | 0 | 0.0021 | 174.4 | 0.197 | 3.00 |
| 29 | 200 | 1/2 | 0 | 0.0021 | 206.6 | 0.233 | 2.54 |
| 27 | 70 | 1 | 0 | 0.0090 | 28.7 | 0.045 | 2.99 |
| 27 | 200 | 1 | 0 | 0.0090 | 36.5 | 0.053 | 2.53 |
| 36 | 70 | 1/2 | 0 | -0.027 | 8.8 | 0.030 | 2.80 |
| 36 | 200 | 1/2 | 0 | -0.027 | 11.2 | 0.036 | 2.37 |
| **32** | **70** | **1** | **0** | **-0.012** | **24.0** | **0.033** | **3.99** |
| **32** | **200** | **1** | **0** | **-0.012** | **30.5** | **0.039** | **3.38** |

TABLE S3. Parameters used to achieve a good agreement between theory and experiment considering that the surface of the LaCoO$_3$ film behaves as a SPM system (zero exchange Heisenberg FM with large effective spin). In bold are indicated the parameters used to compute the temperature dependence of $\Delta\rho_\parallel/\rho_0$ at 9T [black line in Fig. 3(c)] and the FDMR curves at 200 and 70K [see Figs. S3(c) and S3(d)] for the SPM case.

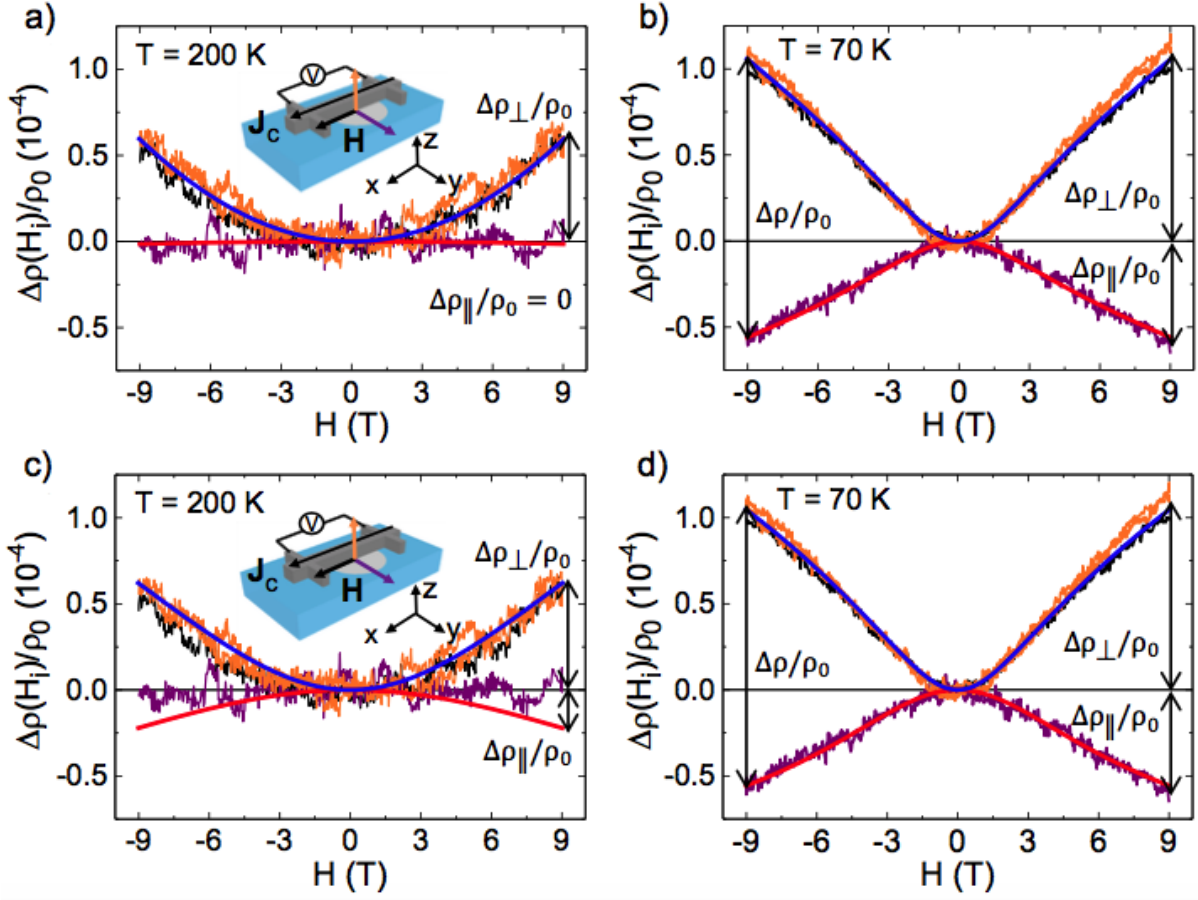

FIG. S6. (a)-(b) Red and blue lines are calculated FDMR curves in LaCoO$_3$/Pt at 200 and 70K using Eqs. (2) and (3) of the main text, together with the experimental data shown in Figs. 2(a) and 2(b), modeling the surface of the LaCoO$_3$ film as a 2D-FM. The parameters used are indicated in Table S1. Excellent agreement between theory and experiment is achieved. (c)-(d) Same but modeling the LaCoO$_3$ surface as a SPM system. The parameters used are indicated in Table S3. In this case, it was impossible to find a set of parameters that were able to reproduce the experimental FDMR curves for both low (70K) and high temperature regimes (200K) at once. For the case represented here, $\Delta\rho_\parallel/\rho_0$ at 200K does not fit well (panel c), ruling out SPM as a plausible magnetic response for the surface of the LaCoO$_3$ film.

Excellent fits were obtained for all considered cases of low-dimensional Heisenberg exchange coupling, but no acceptable fits were obtained for the SPM case for any combination of fitting parameters (see the discrepancy in the calculated $\Delta\rho_\parallel/\rho_0$ –red line– with respect to the experimental curve –purple line– in Fig. S6, which shows one of the best fits), thus excluding this latter scenario. Concretely, we found a better temperature-dependent behavior when considering an exchange coupling between the Co atoms lying between a 1D and a 2D FM case (i.e., for spin ladders, see blue line in Fig. 3(c) of the main text). The results presented in our work are thus a strong indication that low-dimensional FM ordering is taking place at the LaCoO$_3$/Pt interface. However, since the cases of 1D, 1D-ladders and 2D FM ordering are not qualitatively different from each other, we cannot make a firm statement about the exact magnetic arrangement at the LaCoO$_3$ surface. At the moment, we can only discuss different ferromagnetic models and speculate about a possible origin for the magnetic decoupling. We believe that if the C$_{4v}$ symmetry of the LaCoO$_3$ lattice is not broken in LaCoO$_3$ films grown on top of SrTiO$_3$, then the 2D magnetic system should occur. Alternatively, if the C$_{4v}$ is lowered to C$_{2v}$ at the surface, it could well be that the 1D model is realized, but that must be related then to a different scenario of surface reconstruction than what we alluded in Section S5.

## S8. Derivation of Equations (1) and (2) of the main text

Here, we derive the equations used in the main text to interpret the transport measurements in terms of spin-dependent scattering at the Pt/LCO interface. We model the system as a non-magnetic metal in contact with a magnetic interface situated at $z = 0$. Our starting point is the continuity equation for the non-equilibrium spin accumulation in the metal region,

$$\partial_t \mu_{s,j} - \frac{1}{e\nu} \partial_i J_{ij} - \omega_L \varepsilon_{jik} n_i \mu_{s,k} = -\Gamma_{jk} \mu_{s,k} \,. \tag{S1}$$

where $\boldsymbol{\mu}_s$ is the spin bias (e.g. $\mu_{s,z} = \mu_\uparrow - \mu_\downarrow$) which can locally have components in any direction ($j = x, y, z$), the pseudotensor $J_{ij}$ denotes the spin-current flowing in $i$-direction and polarized in $j$-direction, $e$ is the elementary charge, $\nu$ is the density of states per spin species in the metal ($\nu_\uparrow = \nu_\downarrow \equiv \nu$), $\varepsilon_{jik}$ is the total antisymmetric tensor, and repeated indices are implicitly summed over. The last term on the left-hand side describes spin precession, with $\boldsymbol{n} = \boldsymbol{B}/B$ being the unit vector of the magnetic field and $\omega_L = g\mu_B B/\hbar$ the Larmor frequency. An additional contribution to $\omega_L$ occurs due to the exchange field near the LCO surface. We shall return to this point towards the end of this supplemental material.

On the right-hand side in Eq. (S1), $\Gamma_{jk}$ denotes the spin decay tensor which we regard as space dependent. Namely, we assume that the ferromagnetic insulator has a local action on the electrons near the interface over a small depth $b$ into the nonmagnetic metal (no magnetic proximity effect is considered here). We model the spacial dependence of the spin-decay tensor near the interface ($z = 0$) as

$$\Gamma_{jk}(z) = \frac{\theta(z-b)}{\tau_s}\delta_{jk} + \theta(z)\theta(b-z)\Gamma^M_{jk}, \tag{S2}$$

where $\tau_s$ is the spin relaxation time in the nonmagnetic metal (e. g. due to the Elliott-Yafet mechanism [S22]) and $\Gamma^M_{jk}$ is the spin decay tensor, as obtained for electrons traveling in the boundary layer $z \in [0, b]$. We emphasize that we use the boundary layer only as a "trick" to derive a set of boundary conditions. We shall send the thickness $b$ to zero at a later stage in the derivation. For the time being, one can think of the layer $b$ as an auxiliary layer in which the metal and the local moments of the ferromagnetic insulator co-exist in the same region of space.

Readers familiar with the electron spin relaxation on magnetic impurities

To explain the insertion of the $b$-layer in our model, we consider the diffusive motion of an electron in the Pt/LCO system, see Fig. S7 (left panel). The electron moves randomly in the Pt layer, scattering off the Pt/LCO interface once in a while. In the reference frame of the electron, the local moments appear on its trajectory as spikes of interaction of a very short duration at random instances of time. The short duration is related, in particular, to the point-like nature of the coupling between the Co atom and the itinerant electrons in Pt. One can speak of an interaction time,

$$\tau_c = \frac{\max(a_{\text{Co}}, \lambda_F)}{v_F}, \tag{S3}$$

where $a_{\text{Co}}$ is the size of the Co atom and $\lambda_F$ and $v_F$ are, respectively, the Fermi wavelength and Fermi velocity of the itinerant electron. Having scattered away, the electron returns to the interface after a long time, on the oder of the diffusion time, $\tau_D = d_N^2/D$, where $D$ is the diffusion constant and $d_N$ is the thickness of the Pt layer. We assume that, after such a long a time ($t \sim \tau_D$), either the electron spin coherence is lost or the state of the local moments resets, such that the next instance of interaction with the interface can be said to have no memory of the previous one. The short time $\tau_c$ appears then as the correlation time of a spin bath acting on the itinerant electron in the moving reference frame. This bath can be regarded as Markovian [S23] with a good accuracy. The interaction with the local moments can be accompanied by the emission or absorption of a magnon. In the extreme case when $\lambda_F$ is short compared to the distance between local moments on the LCO surface, the electron interacts only with individual local moments on the surface (superradiance effects can be excluded) and the position of these local moments along the $z$-axis is not so crucial, since the electron trajectory is random anyway. This observation leads us to the model system in which the effect of the interface on the spin relaxation is captured by the $b$-layer, see Fig. S7 (right panel). In the system with the $b$-layer, the electron moves diffusively as before, except that the scattering off local moments occurs as in a metal with magnetic impurities. We remark that the ferromagnetic state of the local moments dissolved in the $b$-layer is assumed to be the same as when they were part of the LCO surface, i.e. the magnetic impurities are assumed to be coupled with one another in the same way as when they were part of the LCO surface. Furthermore, we impose the zero-of-all-currents boundary condition at $z = 0$, forbidding the electron to interact with the rest of LCO.

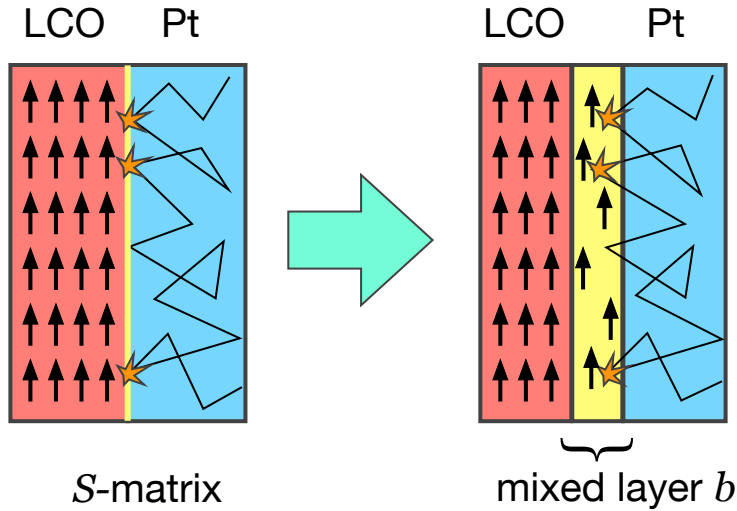

FIG. S7. A sharp interface between a metal (Pt) and a ferromagnetic insulator (LCO) can be described in the language of kinetic equations in terms of a set of boundary conditions, which can, in principle, be obtained from the scattering matrix of the interface. We use an intermediate layer of thickness $b$ as a "trick" to derive these boundary conditions in a straightforward and simple manner by integrating the kinetic equations over the intermediate region and taking subsequently the limit $b \to 0$, see text. Left panel: An electron moving diffusively in Pt and scattering occasionally off the LCO surface interacts from time to time with local moments on the LCO surface, possibly emitting or absorbing a magnon (not shown). Right panel: The Pt/LCO interface with a complicated $S$-matrix is replaced by a metal layer with dilute magnetic impurities. We expect this replacement to be a fair approximation to the original problem when the Fermi wave length in the metal is short compared to the distance between local moments on the surface of the ferromagnetic insulator.

The tensor $\Gamma^M_{jk}$, which describes electron spin relaxation in the $b$-layer, depends on the magnetic state of the interface. For an interface which is magnetically ordered along the B-field, $\Gamma^M_{jk}$ is a uniaxial tensor,

$$\Gamma^M_{ij} = \frac{1}{\tau_\perp}\delta_{ij} + \left(\frac{1}{\tau_\parallel} - \frac{1}{\tau_\perp}\right) n_i n_j, \tag{S4}$$

where $\tau_\parallel$ and $\tau_\perp$ are the longitudinal and transverse spin relaxation times, respectively.

Next, we would like to express the spin relaxation times in terms of microscopic parameters, such as the $s$-$d$ exchange constant, and determine their dependence on the magnetic state of the Co atoms. Before doing so, we need to settle on a model which captures the relevant features of the experiment. It turns out that the presence of static disorder, which is responsible for the diffusive motion of the electron in Pt, is not crucial for the spin relaxation times and can be omitted for the time being from the model. We start with a model Hamiltonian describing the interaction between local moments (of Co atoms) and itinerant electrons in a metal,

$$H = \sum_{\bm{k}s} \varepsilon_{\bm{k}s} c^\dagger_{\bm{k}s} c_{\bm{k}s} + g\mu_B \sum_j \bm{S}_j \cdot \bm{B}$$
$$- J_{sd} \sum_j \bm{S}_j \cdot \bm{s}(\bm{r}_j), \qquad (S5)$$

where $\varepsilon_{\bm{k}s}$ is the energy of an itinerant electron with momentum $\bm{k}$ and spin $s$, $c^\dagger_{\bm{k}s}$ and $c_{\bm{k}s}$ are, respectively, the creation and annihilation operators for the electron, $\bm{S}_j$ is the operator of the local moment (spin $S$) at position $\bm{r}_j$ in the metal, and $\bm{s}(\bm{r}_j)$ is the spin density of itinerant electrons at the site of the local moment. In terms of $c^\dagger_{\bm{k}s}$ and $c_{\bm{k}s}$, the spin density is given by

$$\bm{s}(\bm{r}_j) = \frac{1}{2V} \sum_{\bm{k}\bm{k}'ss'} \bm{\sigma}_{ss'} e^{i(\bm{k}'-\bm{k})\cdot\bm{r}_j} c^\dagger_{\bm{k}s} c_{\bm{k}'s'}, \qquad (S6)$$

where $V$ is the volume of the considred metal layer and $\bm{\sigma} = (\sigma_x, \sigma_y, \sigma_z)$ is a set of Pauli matrices representing the spin 1/2 of the itinerant electron. The second term on the right-hand side in Eq. (S5) is the usual Zeeman interaction in a magnetic field $\bm{B}$, with $g$ being the Landé $g$-factor and $\mu_B$ the Bohr magneton. The last term in Eq. (S5) is the Heisenberg interaction between the local moments and the spin density of the itinerant electrons. The coupling constant $J_{sd}$ arises from the s-d hybridization between the localized d-orbitals of the impurity and the extended s-orbitals of the metal host.

In addition to the terms in Eq. (S5), we take into account the interaction between local moments, modeling it by a Heisenberg exchange term,

$$H_{\text{FM}} = -\sum_{\langle ij \rangle} J_{ij} \bm{S}_i \cdot \bm{S}_j, \qquad (S7)$$

where the sum is taken over pairs of interacting spins without repetition. In plain mean-field approximation (Weiss theory), the interaction between local moments amounts only to a renormalization of the Zeeman term in Eq. (S5),

$$g\mu_B \sum_j \bm{S}_j \cdot \bm{B} \to \sum_j \bm{S}_j \cdot \bm{h}_j, \qquad (S8)$$

where the total exchange field $\bm{h}_j$ acting on the local moment becomes

$$\bm{h}_j = g\mu_{\text{B}} \bm{B} - \sum_j J_{ij} \langle \bm{S}_j \rangle. \qquad (S9)$$

By self-consistency, the average spin is determined by this exchange field,

$$\langle \bm{S}_j \rangle = -S B_S(\beta S h_j) \frac{\bm{h}_j}{h_j}, \qquad (S10)$$

where $B_S(x)$ is the Brillouin function and $\beta = 1/T$. These equations are further simplified in the ferromagnetic case ($J_{ij} > 0$), since all spins have equal averages $\langle \bm{S}_j \rangle = \langle \bm{S} \rangle$ and equal Weiss fields $\bm{h}_j = \bm{h}$, all oriented co-linear with the magnetic field $\bm{B}$. It is convenient to introduce

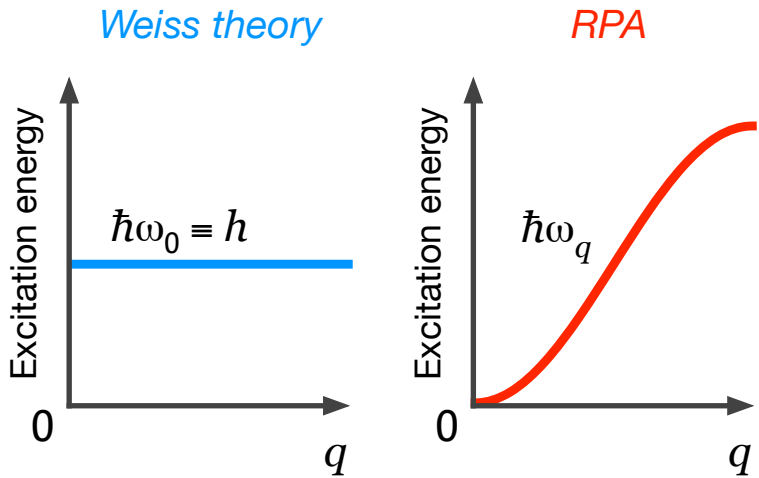

FIG. S8. Excitation spectrum in the case of the Weiss theory and RPA. In the Weiss theory, the elementary excitation on top of the ferromagnetic ground state consists in reducing the absolute value of the $z$-projection of a local moment by unity, which costs an energy equal to the Weiss field $h$. In RPA, a collective excitation (magnon) is possible which also reduces the total ground state spin by unity, but involves a large number of local moments, each of which change their spin states very little relative to their neighbors. Such magnon excitations have a soft spectrum $\omega_q \propto q^2$ at small $q$ and at zero magnetic field, which precludes long-range order in two-dimensions at finite temperatures.

the following notations

$$S_\| = \boldsymbol{n} \cdot \boldsymbol{S},$$
$$\boldsymbol{S}_\perp = \boldsymbol{S} - \boldsymbol{n} S_\| \equiv [\boldsymbol{n} \times [\boldsymbol{S} \times \boldsymbol{n}]], \tag{S11}$$

which represent the spin component along the magnetic field and the remaining transverse spin, respectively. For definiteness, we limit our consideration to the case of positive $g$-factors, $g > 0$, which corresponds to the experiment. The self-consistency equation reduces then to

$$\langle S_\| \rangle = -S B_S \left[ \beta S \left( g\mu_B B - J_0 \langle S_\| \rangle \right) \right], \tag{S12}$$

where $J_0 = \sum_j J_{ij}$. Note that the order parameter $\langle S_\| \rangle$ is negative, meaning that the spins align anti-parallel with the $B$-field for $g > 0$.

Despite the fact that the mean-field theory is usually a satisfactory approximation to the magnetic problem, it cannot be used here to explain the Pt/LCO experiment, because it produces a finite Curie temperature $T_C \sim J$ even in low dimensions (2D and 1D). This prediction is at odds with the experiment which shows no signs of spontaneous or remanent magnetization at $B = 0$. An exact statement is given by the Mermin-Wagner theorem [S25], which states that, for the Heisenberg model in Eq. (S7) with a finite range of interaction (e.g. between nearest neighbors only), no long-range order is possible at arbitrary small but finite temperatures in low dimensions (2D and 1D). The experiment is consistent with what one would expect from such a low-dimensional ferromagnetic model, since it shows a large susceptibility to the magnetic field and absence of the remanent effect. At this point, it becomes clear why the treatment of the magnetic model has to be extended beyond the mean-field approximation. We switch next to the description of the magnetic problem using the random-phase approximation (RPA) for spins [S20], which extends the Weiss theory to account for the width of the magnon band.

The Weiss theory resembles to some extent the Einstein model for phonons, in which all excitations have a single frequency $\omega_0$, forming thus a flat band and making the ground state protected from excitation by the energy gap $\hbar\omega_0$. Acoustic phonons (i.e. phonons with arbitrary small excitation energy) can only be obtained beyond the Einstein model if excitations are allowed to propagate in space, which leads to a band of excitations. This analogy is illustrated in Fig. S8, where the excitation spectrum on top of the ferromagnetic ground state is shown for the Weiss theory (left panel) and RPA (right panel). The only difference to the case of acoustic phonons is that the magnon spectrum is quadratic at small $q$ and not linear as for acoustic phonon. The dependence $\omega_{\bm{q}} \propto q^2$ in the limit $q \to 0$ is a result of the vectorial form of the Heisenberg interaction (S7) which has rotational symmetry and permits neighboring spins to differ very little from each other, creating thus a collective excitation with a small energy cost over a large length scale. Precisely this quadratic dependence is responsible for the absence of long-range ferromagnetic order in low dimensions. The magnon spectrum for the ferromagnetic lattice is given by

$$\hbar\omega_{\bm{q}} = g\mu_{\rm B} B - \left\langle S_{\|} \right\rangle (J_0 - J_{\bm{q}}),$$
$$J_{\bm{q}} = \sum_j J_{ij} e^{i\bm{q}\cdot(\bm{r}_i - \bm{r}_j)}, \tag{S13}$$

where $\left\langle S_{\|} \right\rangle$ is the ferromagnetic order parameter representing the average spin of each local moment. Note that the magnon excitation energy is always positive, since we have $\left\langle S_{\|} \right\rangle < 0$.

The static ferromagnetic correlations are characterized by the moments of the spin operator $S_z$, which can be calculated in RPA from the generating function

$$\Phi(a) = \left\langle e^{-a(\bm{n}\cdot\bm{S})} \right\rangle, \tag{S14}$$

by taking derivatives over the auxiliary parameter $a$, such that $\left\langle S_{\|} \right\rangle = -\Phi'(0)$, $\left\langle S_{\|}^2 \right\rangle = \Phi''(0)$, etc. A method due to Callen [S24] allow to evaluate the generating function

$$\Phi(a) = \frac{m^{2S+1} e^{-aS} - (m+1)^{2S+1} e^{a(S+1)}}{\left[m^{2S+1} - (m+1)^{2S+1}\right]\left[(m+1)e^a - m\right]},$$
$$m = \frac{1}{M} \sum_{\bm{q}} \frac{1}{e^{\beta\hbar\omega_{\bm{q}}} - 1}, \tag{S15}$$

where $M$ is the number of local moments in the lattice. The quantity $m$, which is interpreted as the average number of magnons per site, diverges for the quadratic spectrum in two-dimensions in the thermodynamic limit at zero magnetic field. As a result, the ferromagnetic ground state is destroyed by the soft mode of excitations at arbitrarily small but finite temperatures. This important feature follows directly from the self-consistency condition, which is obtained from the generating function to be

$$\left\langle S_{\|} \right\rangle = \frac{(S-m)(m+1)^{2S+1} + (S+m+1)^{2S+1}}{m^{2S+1} - (m+1)^{2S+1}}. \tag{S16}$$

Namely, the algebraic expression on the right-hand side here vanishes in the limit $m \to \infty$, leading to $\left\langle S_{\|} \right\rangle = 0$, consistent with the Mermin-Wagner theorem [S25]. In practice, the order parameter $\left\langle S_{\|} \right\rangle$ is found by solving simultaneously Eqs. (S13) and (S16) and using the definition of $m$ in Eq. (S15).

The RPA allows also to evaluate certain correlators describing the propagation of the spin excitation in space and time, which then allow to introduce the magnetic correlation length $\xi$. The physical picture arising for the magnetism of the ferromagnetic Heisenberg model in 2D is as follows. One can group local moments into a cluster of correlated spins gathered from an area $\sim \xi^2$. The correlation length depends exponentially on the temperature $T$ [S26],

$$\xi \sim e^{\frac{\pi T_W}{T}}, \tag{S17}$$

where $T_W$ is a characteristic scale given by the Curie temperature occurring in the Weiss theory. Thus, one can regard the ferromagnet as consisting of independent clusters of local moments, each with a total spin $S_{\text{cluster}} \simeq S\xi^2$. This resembles to some extent a superparamagnet: a collection of large spins which get relatively easily aligned in a magnetic field (hence large susceptibility), but at the same time do not undergo a phase transition. It is, therefore, not accidental that the experimental data for Pt/LCO can qualitatively be explained by a superparamagnetic behavior of the LCO surface. However, the fits cannot be made accurate when the temperature dependence of the measured resistivity correction is considered (see main text). With this remark we conclude our qualitative description of RPA. For further details and for a rigorous derivation of Eqs. (S13)–(S16), we refer the reader to the book by Majlis [S20].

We turn now to calculating the spin relaxation times in the Born-Markov approximation [S23]. We consider first an electron with momentum $\boldsymbol{k}$ and spin $s$ interacting with a single magnetic impurity and flipping its spin. According to Fermi's golden rule, the microscopic transition rate is

$$W_{\bar{s},s}(\boldsymbol{k}', \boldsymbol{k}) = \int_{-\infty}^{+\infty} \frac{dt}{\hbar^2} e^{-\frac{it}{\hbar}(\varepsilon_{\boldsymbol{k}'\bar{s}} - \varepsilon_{\boldsymbol{k}s})} \overline{\langle \boldsymbol{k}s | U_{\text{int}}(t) | \boldsymbol{k}'\bar{s} \rangle \langle \boldsymbol{k}'\bar{s} | U_{\text{int}} | \boldsymbol{k}s \rangle}, \tag{S18}$$

where the initial and final electronic states are

$$|\boldsymbol{k}s\rangle = c^\dagger_{\boldsymbol{k}s} |0\rangle,$$
$$|\boldsymbol{k}'\bar{s}\rangle = c^\dagger_{\boldsymbol{k}'\bar{s}} |0\rangle, \tag{S19}$$

with $\bar{s} = -s$ denoting the flipped spin. The perturbation entering in Eq. (S18) represents the interaction of the itinerant electron with a single magnetic impurity, which we can assume without loss of generality to reside at $\boldsymbol{r} = 0$,

$$U_{\text{int}} = -J_{sd} \boldsymbol{S} \cdot \boldsymbol{s}(0), \tag{S20}$$

see also the last term in Eq. (S5). The time dependence of the perturbation in Eq. (S18) is taken in the interaction picture with the Hamiltonian of the "bath". In our model, this is the Hamiltonian of the local moments, which consists of the Zeeman interaction, given by the second term in Eq. (S5), and the Heisenberg exchange in Eq. (S7). Furthermore, in Eq. (S18), we average over the state of the magnetic system at thermal equilibrium, which is denoted by the bar over the product of the two matrix elements, i.e. $\overline{\langle i | U(t) | f \rangle \langle f | U | i \rangle}$. Later, we shall denote the same average by $\langle \ldots \rangle$, after taking the matrix elements between the electronic states.

In the Born approximation, we take into account the contributions of different local moments in an incoherent fashion, by multiplying the rate in Eq. (S18) by the number of local moments in the volume of the $b$-layer. It is convenient to make use of the 2D concentration of the local moments on the surface to express the 3D concentration in the $b$-layer as

$$n_{imp}^{3D} = \frac{n_{imp}^{2D}}{b}. \tag{S21}$$

We obtain then from Eq. (S18) the rate for the electron $|\boldsymbol{k}s\rangle$ to flip its spin while scattering off magnetic impurities in the $b$-layer to be

$$W_{\downarrow\uparrow} = \frac{n_{imp}^{3D} J_{sd}^2}{4\hbar^2 V} \int_{-\infty}^{\infty} dt e^{\frac{it}{\hbar}(\varepsilon_{\boldsymbol{k}\uparrow} - \varepsilon_{\boldsymbol{k}'\downarrow})} \langle S_-(t) S_+ \rangle,$$
$$W_{\uparrow\downarrow} = \frac{n_{imp}^{3D} J_{sd}^2}{4\hbar^2 V} \int_{-\infty}^{\infty} dt e^{\frac{it}{\hbar}(\varepsilon_{\boldsymbol{k}\downarrow} - \varepsilon_{\boldsymbol{k}'\uparrow})} \langle S_+(t) S_- \rangle, \tag{S22}$$

where $S_+ = \langle \downarrow | \boldsymbol{\sigma} | \uparrow \rangle \cdot \boldsymbol{S}$ and $S_- = \langle \uparrow | \boldsymbol{\sigma} | \downarrow \rangle \cdot \boldsymbol{S}$ are the ladder operators for the local moment with the quantization axis chosen along $\boldsymbol{n}$. The ladder operators satisfy the commutation relation [S27] $[S_+, S_-] = 2S_\parallel$, where $S_\parallel$ is defined in Eq. (S11). The spin-up and spin-down states for the itinerant electron are chosen as eigenvectors of the matrix $\boldsymbol{n} \cdot \boldsymbol{\sigma}$,

$$|\uparrow\rangle = \begin{pmatrix} \cos\frac{\theta}{2} \\ e^{i\phi}\sin\frac{\theta}{2} \end{pmatrix}, \qquad |\downarrow\rangle = \begin{pmatrix} -e^{-i\phi}\sin\frac{\theta}{2} \\ \cos\frac{\theta}{2} \end{pmatrix}, \qquad (S23)$$

where $\theta$ and $\phi$ are the polar angles of the unit vector $\boldsymbol{n} = \{\cos\phi\sin\theta, \sin\phi\sin\theta, \cos\theta\}$. Clearly, we do not have to perform the calculation in this spin basis. In practice, we set the spin quantization axis to $\boldsymbol{n} = \{0, 0, 1\}$ and calculate the matrix elements in the usual spin basis of the $\sigma_z$ Pauli matrix and similarly we proceed with the magnetic problem, since it is rotationally invariant. However, to avoid confusion with the notations, we are obliged to use here $S_\parallel$ and $\boldsymbol{S}_\perp$ to denote the longitudinal and transverse components of the spin.

With the rates in Eq. (S22), we are in position to describe the longitudinal spin relaxation. We introduce two distribution functions $f_\uparrow(E)$ and $f_\downarrow(E)$ to describe the out-of-equilibrium spin accumulation in the itinerant electron system. The number of spin-up and spin-down electrons per unit volume is given by

$$N_{\uparrow/\downarrow} = \frac{1}{V}\sum_{\boldsymbol{k}} f_{\uparrow/\downarrow}(\varepsilon_{\boldsymbol{k}\uparrow/\downarrow}) \equiv \int dE\, \nu_{\uparrow/\downarrow}(E) f_{\uparrow/\downarrow}(E), \qquad (S24)$$

where $\nu_\uparrow(E)$ and $\nu_\downarrow(E)$ are the densities of states of the two spin species. Because of the Zeeman and exchange fields, $\nu_\uparrow(E)$ and $\nu_\downarrow(E)$ differ, in general, from each other. In a single-band model, this difference can be expressed via a single function $\nu(E)$ shifted in energy by the spin splitting,

$$\nu_\uparrow(E) = \nu\left(E - \hbar\omega_L/2\right), \quad \nu_\downarrow(E) = \nu\left(E + \hbar\omega_L/2\right),$$

where $\nu(E)$ is the density of states in the absence of Zeeman and exchange fields and $\hbar\omega_L$ is the total splitting due to these fields. Because of the difference between $\nu_\uparrow(E)$ and $\nu_\downarrow(E)$, a spin polarization is present already in equilibrium ($N_\uparrow \neq N_\downarrow$). However, we are interested in the out-of-equilibrium spin polarization, which appears when the distribution functions $f_\uparrow(E)$ and $f_\downarrow(E)$ differ from each other in a particular way. To make this statement rigorous, we define the out-of-equilibrium spin polarization as

$$P = \delta N_\uparrow - \delta N_\downarrow,$$
$$\delta N_{\uparrow/\downarrow} = N_{\uparrow/\downarrow} - N_{\uparrow/\downarrow}^{\text{eq}}, \qquad (S25)$$

where $N_{\uparrow/\downarrow}^{\text{eq}}$ are the equilibrium concentrations, obtained after replacing $f_{\uparrow/\downarrow}(E)$ in Eq. (S24) by the Fermi-Dirac distribution function

$$f(E) = \frac{1}{e^{\beta(E-\mu)} + 1}, \qquad (S26)$$

where $\mu$ is the electrochemical potential.

We write down balance equations for the spin-up and spin-down electron concentrations,

$$\frac{\partial N_\uparrow}{\partial t} = -\frac{1}{V} \sum_{\bm{kk'}} W_{\downarrow\uparrow}(\bm{k'},\bm{k}) f_\uparrow(\varepsilon_{\bm{k}\uparrow})[1 - f_\downarrow(\varepsilon_{\bm{k'}\downarrow})]$$

$$+ \frac{1}{V} \sum_{\bm{kk'}} W_{\uparrow\downarrow}(\bm{k},\bm{k'}) f_\downarrow(\varepsilon_{\bm{k'}\downarrow})[1 - f_\uparrow(\varepsilon_{\bm{k}\uparrow})],$$

$$\frac{\partial N_\downarrow}{\partial t} = -\frac{1}{V} \sum_{\bm{kk'}} W_{\uparrow\downarrow}(\bm{k'},\bm{k}) f_\downarrow(\varepsilon_{\bm{k}\downarrow})[1 - f_\uparrow(\varepsilon_{\bm{k'}\uparrow})]$$

$$+ \frac{1}{V} \sum_{\bm{kk'}} W_{\downarrow\uparrow}(\bm{k},\bm{k'}) f_\uparrow(\varepsilon_{\bm{k'}\uparrow})[1 - f_\downarrow(\varepsilon_{\bm{k}\downarrow})],$$

(S27)

where the factors $f(E)[1-f(E')]$ account for the fact that the carriers are fermions. The rates in Eq. (S22) obey the detailed balance principle

$$W_{\downarrow\uparrow}(\bm{k'},\bm{k}) = e^{\beta(\varepsilon_{\bm{k}\uparrow} - \varepsilon_{\bm{k'}\downarrow})} W_{\uparrow\downarrow}(\bm{k},\bm{k'}). \quad (S28)$$

This relation alone suffices to show that the right-hand side in Eq. (S27) vanishes at equilibrium. Indeed, replacing $f_{\uparrow/\downarrow}(E)$ by the Fermi-Dirac distribution and using the identity

$$f(\varepsilon_\uparrow)[1 - f(\varepsilon_\downarrow)] = e^{\beta(\varepsilon_\downarrow - \varepsilon_\uparrow)} f(\varepsilon_\downarrow)[1 - f(\varepsilon_\uparrow)], \quad (S29)$$

we find from Eq. (S27) that $\partial N_\uparrow/\partial t = 0$ and $\partial N_\downarrow/\partial t = 0$, as expected for equilibrium.

Although Eq. (S27) were written for the whole sample, they can also be applied to a small but still sufficiently macroscopic volume $V$ of the sample, provided we add additional terms accounting for the influx of carriers through the surface of that volume. Thus, we replace the left-hand side in Eq. (S27) as

$$\frac{\partial N_{\uparrow/\downarrow}}{\partial t} \to \frac{\partial N_{\uparrow/\downarrow}}{\partial t} + \frac{1}{V} \iint d\bm{A} \cdot \bm{j}_{\uparrow/\downarrow} \quad (S30)$$

where $\bm{j}_{\uparrow/\downarrow}$ is the particle current of the given spin species and $d\bm{A}$ is the element of the surface area for the volume $V$. Going to the limit of a small volume as compared to the characteristic scale over which $\bm{j}_{\uparrow/\downarrow}$ changes, one recovers the usual diffusion term

$$\frac{\partial N_{\uparrow/\downarrow}}{\partial t} \to \frac{\partial N_{\uparrow/\downarrow}}{\partial t} + \operatorname{div} \bm{j}_{\uparrow/\downarrow}. \quad (S31)$$

At equilibrium, we then have $\partial N_{\uparrow/\downarrow}/\partial t = 0$ and $\bm{j}_{\uparrow/\downarrow} = 0$.

Next, we consider the out-of-equilibrium situation illustrated in Fig. S9. We assume that the equilibration is fast within each spin species separately and the terms present in Eq. (S27) describe only the "bottle neck" of the relaxation to thermal equilibrium. In this case, $f_\uparrow(E)$ and $f_\downarrow(E)$ can both be assumed to be Fermi-Dirac distributions, except that each distribution is shifted to the electrochemical potential of its own spin species. We approximate

$$f_\uparrow(E) \approx \frac{1}{e^{\beta(E-\mu_\uparrow)} + 1}, \quad f_\downarrow(E) \approx \frac{1}{e^{\beta(E-\mu_\downarrow)} + 1}, \quad (S32)$$

where $\mu_\uparrow$ and $\mu_\downarrow$ are the spin-up and spin-down electrochemical potentials, respectively. Introducing,

$$\mu_{s,z} = \mu_\uparrow - \mu_\downarrow, \quad (S33)$$

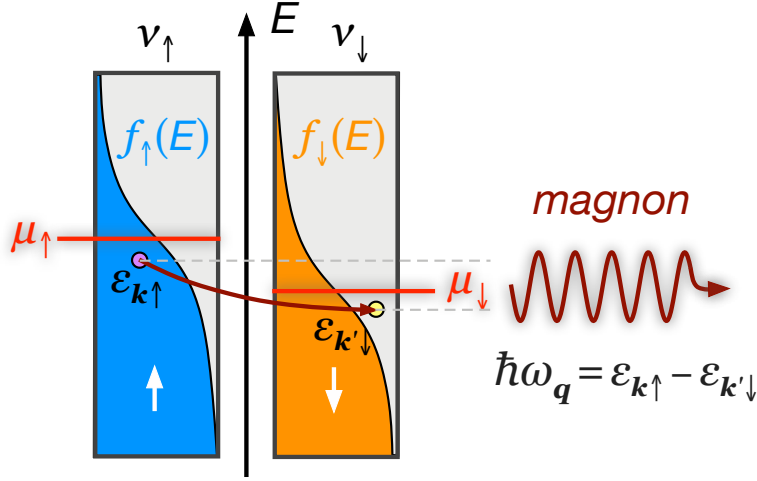

FIG. S9. An out-of-equilibrium spin polarization is achieved by two different occupations of the spin-up ($\nu_\uparrow$) and spin-down ($\nu_\downarrow$) densities of states. We assume that equilibration is fast within each spin species, such that they are populated according to the Fermi-Dirac distribution function, which is shifted to two different electrochemical potentials $\mu_\uparrow$ and $\mu_\downarrow$ for the two spins, respectively. The spin bias $\mu_\uparrow - \mu_\downarrow$ relaxes due to the electron scattering with spin flip, accompanied by the emission/absorption of a magnon. The electron scattering obeys the Pauli exclusion principle, which is accounted in the rate by occupation factors, such as $f(\varepsilon_{\boldsymbol{k}\uparrow} - \mu_\uparrow)\left[1 - f(\varepsilon_{\boldsymbol{k}'\downarrow} - \mu_\downarrow)\right]$ for the process depicted here.

it is straightforward to rewrite Eq. (S27) as

$$\begin{aligned}\frac{\partial N_\uparrow}{\partial t} &= -\frac{1}{V}\sum_{\boldsymbol{k}\boldsymbol{k}'} W_{\downarrow\uparrow}\left(\boldsymbol{k}', \boldsymbol{k}\right) f_\uparrow\left(\varepsilon_{\boldsymbol{k}\uparrow}\right)\left[1 - f_\downarrow\left(\varepsilon_{\boldsymbol{k}'\downarrow}\right)\right] \\ &\quad \times \left(1 - e^{-\beta\mu_{s,z}}\right), \\ \frac{\partial N_\downarrow}{\partial t} &= -\frac{1}{V}\sum_{\boldsymbol{k}\boldsymbol{k}'} W_{\uparrow\downarrow}\left(\boldsymbol{k}', \boldsymbol{k}\right) f_\downarrow\left(\varepsilon_{\boldsymbol{k}\downarrow}\right)\left[1 - f_\uparrow\left(\varepsilon_{\boldsymbol{k}'\uparrow}\right)\right] \\ &\quad \times \left(1 - e^{\beta\mu_{s,z}}\right), \end{aligned} \quad (S34)$$

where we omitted the terms div $\boldsymbol{j}_{\uparrow/\downarrow}$, which can always be added at the final stage as discussed above. The total charge is conserved both in the original equation (S27) and in the approximate equation (S34),

$$\frac{\partial\left(N_\uparrow + N_\downarrow\right)}{\partial t} + \mathrm{div}\left(\boldsymbol{j}_\uparrow + \boldsymbol{j}_\downarrow\right) = 0. \quad (S35)$$

In linear response, the deviation from equilibrium is weak and we take the limit $\mu_{s,z} \to 0$ in Eq. (S34). In addition, we focus on a range of energies on the order of the temperature near the Fermi surface and assume that the density of states is constant in this range and equal for both spin species,

$$\nu_\uparrow \approx \nu, \qquad \nu_\downarrow \approx \nu. \quad (S36)$$

It is easy then to see that

$$\frac{\partial\left(N_\uparrow - N_\downarrow\right)}{\partial t} = \nu\frac{\partial\left(\mu_\uparrow - \mu_\downarrow\right)}{\partial t}, \quad (S37)$$

which together with Eqs. (S31) and (S34) leads to the kinetic equation for the longitudinal spin bias $\mu_{s,z}$,

$$\frac{\partial\mu_{s,z}}{\partial t} + \frac{1}{\nu}\,\mathrm{div}\left(\boldsymbol{j}_\uparrow - \boldsymbol{j}_\downarrow\right) = -\frac{\mu_{s,z}}{\tau_\parallel}, \quad (S38)$$

where $\tau_\parallel$ is the longitudinal spin relaxation time. We obtain
$$\frac{1}{\tau_\parallel} = w_{\downarrow\uparrow} + w_{\uparrow\downarrow}, \tag{S39}$$
where
$$w_{\downarrow\uparrow} = \frac{1}{V\nu T} \sum_{\boldsymbol{k}\boldsymbol{k}'} W_{\downarrow\uparrow}(\boldsymbol{k}', \boldsymbol{k}) f(\varepsilon_{\boldsymbol{k}\uparrow}) [1 - f(\varepsilon_{\boldsymbol{k}'\downarrow})],$$
$$w_{\uparrow\downarrow} = \frac{1}{V\nu T} \sum_{\boldsymbol{k}\boldsymbol{k}'} W_{\uparrow\downarrow}(\boldsymbol{k}', \boldsymbol{k}) f(\varepsilon_{\boldsymbol{k}\downarrow}) [1 - f(\varepsilon_{\boldsymbol{k}'\uparrow})]. \tag{S40}$$

We remark that Eq. (S39) resembles the formula for the longitudinal spin-relaxation time $T_1$ of a spin qubit,
$$\frac{1}{T_1} = w_{\downarrow\uparrow} + w_{\uparrow\downarrow}, \tag{S41}$$
where $w_{\bar{s}s}$ is the rate to go from spin $s$ to spin $\bar{s} = -s$. In the case of Eq. (S39), $w_{\bar{s}s}$ represents an average rate for the electrons to flip spin by emitting a magnon. The microscopic process with the rate $W_{\downarrow\uparrow}(\boldsymbol{k}', \boldsymbol{k})$ is illustrated in Fig. S9. In order to obtain an effective rate, the microscopic rate needs to be summed over the final states and averaged over the initial state. This is done for $w_{\downarrow\uparrow}$ in Eq. (S40) as follows. The sum over $\boldsymbol{k}'$ with the occupation factor $[1 - f(\varepsilon_{\boldsymbol{k}'\downarrow})]$ performs the summation over the final states (sum over all holes in the Fermi sea), whereas the average over the initial state is carried out by
$$\frac{1}{V\nu T} \sum_{\boldsymbol{k}} f(\varepsilon_{\boldsymbol{k}\uparrow}) \ldots, \tag{S42}$$
which can be interpreted as a sum over all electrons in the Fermi sea, divided by the number of states at the Fermi surface in the energy window $T$. Of course, most of the electrons are not able to make a transition, due to energy conservation and Pauli statistics, and the contribution to the sum in Eq. (S42) comes mostly from the electrons at the Fermi surface in the energy window on the order of $T$.

Using Eqs. (S28) and (S29), it is straightforward to show that the two effective rates in Eq. (S40) are equal to each other,
$$w_{\downarrow\uparrow} = w_{\uparrow\downarrow}. \tag{S43}$$
This is a direct consequence of our assumption about the density of states in Eq. (S36). Namely, what matters here is that we assumed that the two spin species have identical density of states, $\nu_\uparrow(E) = \nu_\downarrow(E)$. This is not the case for the spin qubit and the rates $w_{\downarrow\uparrow}$ and $w_{\uparrow\downarrow}$ differ from each other strongly at low tempratures. Using the property in Eq. (S43), one can rewrite Eq. (S39) as
$$\frac{1}{\tau_\parallel} = 2w_{\downarrow\uparrow} = 2w_{\uparrow\downarrow}. \tag{S44}$$

Substituting Eq. (S22) into Eq. (S40) and going from summation to integration, we obtain

$$w_{\downarrow\uparrow} = \frac{\pi n_{imp}^{3D} \nu J_{sd}^2}{2\hbar T} \int_{-\infty}^{\infty} dE f(E)$$
$$\times \int_{-\infty}^{\infty} d\omega \left[1 - f(E - \hbar\omega)\right] D_{-+}(\omega),$$
$$w_{\uparrow\downarrow} = \frac{\pi n_{imp}^{3D} \nu J_{sd}^2}{2\hbar T} \int_{-\infty}^{\infty} dE f(E)$$
$$\times \int_{-\infty}^{\infty} d\omega \left[1 - f(E - \hbar\omega)\right] D_{+-}(\omega), \quad (S45)$$

where

$$D_{-+}(\omega) = \frac{1}{2\pi} \int_{-\infty}^{\infty} dt e^{i\omega t} \langle S_-(t) S_+ \rangle,$$
$$D_{+-}(\omega) = \frac{1}{2\pi} \int_{-\infty}^{\infty} dt e^{i\omega t} \langle S_+(t) S_- \rangle. \quad (S46)$$

are the dynamical structure factors [S28] of the ladder operators. By their definition, the dynamical structure factors obey [c.f. Eq. (S28)]

$$D_{-+}(\omega) = e^{\beta\hbar\omega} D_{+-}(-\omega), \quad (S47)$$

which can be used as a direct way to verify that the expressions for the rates in Eq. (S45) satisfy Eq. (S43).

The dynamical structure factors can be conveniently related to the Green function used in RPA for the magnetic system [S20]. By the virtue of the fluctuation-dissipation theorem [S28], we have

$$D_{-+}(\omega) = -\frac{\hbar}{\pi} \left[1 + n_B(\omega)\right] \Im\left[\chi_{-+}(\omega)\right], \quad (S48)$$

where $\Im[\ldots]$ stands for the imaginary part and $n_B(\omega)$ is the Bose-Einstein distribution function

$$n_B(\omega) = \frac{1}{e^{\beta\hbar\omega} - 1}. \quad (S49)$$

Introducing the retarded linear response function

$$\chi_{-+}(t) = -\frac{i}{\hbar} \theta(t) \langle [S_-(t), S_+] \rangle, \quad (S50)$$

we have

$$\chi_{-+}(\omega) = \int_{-\infty}^{\infty} dt e^{i\omega t} \chi_{-+}(t)$$
$$= -\frac{i}{\hbar} \int_0^{\infty} dt e^{i(\omega + i0)t} \langle [S_-(t), S_+] \rangle. \quad (S51)$$

In RPA, $\chi_{-+}(\omega)$ can be decomposed into a sum over poles, each of which represents a magnon excitation. We leave out the details of the derivation and refer the reader to Refs. S20 and S26. One obtains

$$\chi_{-+}(\omega) = -\frac{1}{M\hbar} \sum_{\boldsymbol{q}} \frac{2 \langle S_\parallel \rangle}{\omega - \omega_{\boldsymbol{q}} + i0}, \quad (S52)$$

where $\omega_{\boldsymbol{q}}$ is the magnon spectrum in Eq. (S13). Substituting Eq. (S52) into Eq. (S48), we obtain the spectral decomposition of the dynamical structure factor

$$D_{-+}(\omega) = -\frac{2\langle S_{\|}\rangle}{M} \sum_{\bm{q}} [1 + n_B(\omega_{\bm{q}})] \delta(\omega - \omega_{\bm{q}}). \tag{S53}$$

The physical meaning of this expression is as follows. Since, in our notations, the ferromagnetic state has the spins mostly oriented anti-parallel to the magnetic field, the correlator $\langle S_-(t)S_+\rangle$, and hence $D_{-+}(\omega)$, describes the creation of magnon excitations. Each magnon adds to the ferromagnetic state of a positive spin projection of $\Delta S_{\|} = 1$. A magnon excitation is accomplished by raising the energy of the magnetic system by $\hbar\omega_{\bm{q}}$ and lowering the absolute value of the total spin by 1, deviating thus the system further from the ground state. The factor $1 + n_B(\omega_{\bm{q}})$ in Eq. (S53) describes the combination of spontaneous and stimulated emissions, typical for bosonic excitations. Owing to the spontaneous emission, magnons can be created even at zero temperature, when the magnetic system is in its ferromagnetic ground state and no magnons are available in the system ($n_B \equiv 0$).

Going along the same lines of derivation as above, we obtain for the other correlator

$$D_{+-}(\omega) = -\frac{2\langle S_{\|}\rangle}{M} \sum_{\bm{q}} n_B(\omega_{\bm{q}}) \delta(\omega + \omega_{\bm{q}}). \tag{S54}$$

The difference here with respect to Eq. (S53) is that $D_{+-}(\omega)$ describes a change of spin by $\Delta S_{\|} = -1$, which is interpreted as an absorption of a magnon from the ferromagnet; hence, the opposite frequency in the $\delta$-function and the presence of the stimulated emission factor $n_B(\omega_{\bm{q}})$ without the possibility of spontaneous emission. This means that magnon absorption is impossible at zero temperature ($n_B \equiv 0$). Clearly, no energy can then be taken from the ferromagnet (because it is in the ground state) and its spin is already maximally polarized.

Substituting Eqs. (S53) and (S54) into Eq. (S45), we obtain

$$\begin{aligned}w_{\downarrow\uparrow} &= \frac{\pi n_{imp}^{3D} \nu J_{sd}^2}{\hbar T} \int_{-\infty}^{\infty} dE f(E) \\ &\times \frac{|\langle S_{\|}\rangle|}{M} \sum_{\bm{q}} [1 - f(E - \hbar\omega_{\bm{q}})][1 + n_B(\omega_{\bm{q}})], \\ w_{\uparrow\downarrow} &= \frac{\pi n_{imp}^{3D} \nu J_{sd}^2}{\hbar T} \int_{-\infty}^{\infty} dE f(E) \\ &\times \frac{|\langle S_{\|}\rangle|}{M} \sum_{\bm{q}} [1 - f(E + \hbar\omega_{\bm{q}})] n_B(\omega_{\bm{q}}).\end{aligned} \tag{S55}$$

In these expressions, one can still trace the microscopic scattering process accompanied by the emission (for $w_{\downarrow\uparrow}$) and absorption (for $w_{\uparrow\downarrow}$) of a magnon, as described above. The presence of the order parameter $\langle S_{\|}\rangle$ as a prefactor in Eqs. (S52)–(S55) has a special meaning. It plays the role of a $Z$-factor in the Green function and therefore serves as a measure of strength (or efficiency) of coupling between a magnon and a local moment. In the ground state, when $\langle S_{\|}\rangle = -S$, this factor shows that the larger the value of $S$ the easier is to excite a magnon by coupling to a local moment. Since we are using the RPA, the magnon appears in our expressions as a well-defined excitation without a line-width [S29]. The temperature dependence of $\langle S_{\|}\rangle$ can be interpreted as a renormalization effect of the self-consistent field. At finite temperatures, the value of $\langle S_{\|}\rangle$ is determined by $n_B(\omega_{\bm{q}})$ via the self-consistency loop of Eqs. (S13), (S15), and (S16). In particular, the quantity $m$ in Eq. (S15) can be written as

$$m = \frac{1}{M} \sum_{\boldsymbol{q}} n_B(\omega_{\boldsymbol{q}}). \tag{S56}$$

Subsequently, $m$ determines $\langle S_\parallel \rangle$ by Eq. (S16). These relations show that the Bose-Einstein distribution function appears in Eqs. (S53) and (S54) implicitly via the quantity $\langle S_\parallel \rangle$, which may create certain ambiguities in the way of writing those equations and in their interpretation.

Next, we integrate over the energy in Eq. (S55) using

$$\int_{-\infty}^{\infty} dE f(E) [1 - f(E - \hbar\omega_{\boldsymbol{q}})] = \hbar\omega_{\boldsymbol{q}} n_B(\omega_{\boldsymbol{q}}),$$

$$\int_{-\infty}^{\infty} dE f(E) [1 - f(E + \hbar\omega_{\boldsymbol{q}})] = \hbar\omega_{\boldsymbol{q}} [1 + n_B(\omega_{\boldsymbol{q}})].$$

We obtain $w_{\downarrow\uparrow} = w_{\uparrow\downarrow}$, with

$$w_{\downarrow\uparrow} = \frac{\pi n_{imp}^{3D} \nu J_{sd}^2 |\langle S_\parallel \rangle|}{TM} \sum_{\boldsymbol{q}} \omega_{\boldsymbol{q}} n_B(\omega_{\boldsymbol{q}}) [1 + n_B(\omega_{\boldsymbol{q}})], \tag{S57}$$

which can also be written as

$$w_{\downarrow\uparrow} = \frac{\pi n_{imp}^{3D} \nu J_{sd}^2}{\hbar} \langle S_\parallel \rangle \beta \frac{\partial m}{\partial \beta}. \tag{S58}$$

Together with Eq. (S39) this expression for the rates gives the longitudinal spin relaxation time $\tau_\parallel$ within the RPA for the magnetic system. We remark that Eq. (S57) can be given many equivalent forms by exploiting the self-consistency equation as an identity.

In the extreme case of a 2D Heisenberg ferromagnet $\langle S_\parallel \rangle$ vanishes at $B = 0$. However, this does not mean that the rates $w_{\downarrow\uparrow}$ and $w_{\uparrow\downarrow}$ vanish as well. As we already mentioned above, the average magnon occupation $m$ becomes infinite at this point. It is easy to see that the product $\langle S_\parallel \rangle n_B(\omega_{\boldsymbol{q}})$ tends to a constant in the limit $\langle S_\parallel \rangle \to 0$,

$$-\langle S_\parallel \rangle n_B(\omega_{\boldsymbol{q}}) \approx \frac{T}{(J_0 - J_{\boldsymbol{q}}) - g\mu_B B / \langle S_\parallel \rangle}, \tag{S59}$$

where the ratio $-g\mu_B B / \langle S_\parallel \rangle$ assumes a constant value in the limit $B \to 0$. This value is inversely proportional to the initial static susceptibility $\chi = -g\mu_B \lim_{B \to 0} \langle S_\parallel \rangle / B$ per magnetic impurity. The deviation of the susceptibility from the Curie-Weiss law acquires in RPA theory the form of the following equation for $\chi$ [S26, S20]

$$\frac{1}{V} \sum_{\boldsymbol{q}} \left[ \frac{(J_0 - J_{\boldsymbol{q}})}{(g\mu_B)^2} + \frac{1}{\chi} \right]^{-1} = \frac{C}{T}, \tag{S60}$$

where $C = \frac{1}{3}(g\mu_B)^2 S(S+1) n_{imp}^{3D}$ is the Curie constant, with $n_{imp}^{3D} = M/V$ being the concentration of magnetic impurities. Due to the exchange coupling $J_{ij}$ which enters in Eq. (S60) via $J_0$ and $J_{\boldsymbol{q}}$, the susceptibility $\chi$ is enhanced as compared to the paramagnetic Curie-Weiss law. Taking the limit $B \to 0$ in Eq. (S57) with the help of Eqs. (S59) and (S60), we obtain

$$w_{\downarrow\uparrow} = w_{\uparrow\downarrow} = \frac{\pi n_{imp}^{3D} \nu J_{sd}^2 S(S+1)}{3\hbar}, \tag{S61}$$

which is identical to the spin-flip rates for electron scattering off paramagnetic impurities. Thus,

despite having substantial static correlations building up with lowering the temperature over an exponentially large correlation length $\xi$, see Eq. (S17), in our calculation the electron "sees" only a small portion of the interface and interacts at most with one local moment. The absence of a bandwidth in the magnon band in the limit $B \to 0$ renders the scattering identical to the paramagnetic case. It is easy to verify that, without a bandwidth, the RPA results reproduce exactly the paramagnetic ones. In particular, by setting $\langle S_\| \rangle = 0$ in the spectrum $\hbar\omega_{\boldsymbol{q}}$ in Eq. (S13) and using the expression for $m$ in Eq. (S15) with such a "paramagnetic" spectrum, one obtains for the spin average in Eq. (S16) the result expected in the paramagnetic case: $\langle S_\| \rangle = -SB_S\left(\beta Sg\mu_\mathrm{B}B\right)$, which coincides also with Eq. (S12), if $J_0$ is set to zero therein. By going beyond the Born approximation, one is expected to obtain corrections to paramagnetic result in Eq. (S61), but we do not pursue it here.

Next, we consider the high-temperature limit in which the order parameter $\langle S_\| \rangle$ is sufficiently small such that the energy $\hbar\omega_{\boldsymbol{q}}$ of the characteristic magnon emitted/absorbed during scattering is small compared to the temperature. This is the limit of the so-called "frozen" magnon field, which acts as a static potential, being unable to take or give considerable energy to the electron in each scattering event. The electron scattering becomes then essentially elastic. We would like to obtain general results which are not specific to RPA. To do so we go back to Eq. (S45) and expand the fermionic factor $[1 - f(E - \hbar\omega)]$ under the integral for small inelastic energies $\hbar\omega/T \ll 1$. On technical reasons, it is convenient to use $p = 1 - e^{-\beta\hbar\omega}$ as a small quantity during the expansion. This allows to rearrange the series right away in such as way that no re-summation is required. We thus expand the fermionic factor in powers of $p$ as follows [S32]

$$1 - f(E - \hbar\omega) = 1 - f(E) - \sum_{k=1}^{\infty} f(E)\left[1 - f(E)\right]^k p^k \qquad (S62)$$

Keeping the terms up to the first order of $p$, we obtain

$$w_{\downarrow\uparrow} = w_{\uparrow\downarrow} = \frac{\pi n_{imp}^{3D}\nu J_{sd}^2}{2\hbar}\left\langle \boldsymbol{S}_\perp^2 \right\rangle, \qquad (S63)$$

which shows that, in the elastic-collision limit, the information about the magnetic system enters in the spin-flip rates only through the variance of the transverse spin fluctuations. To obtain Eq. (S63), we used several general properties of the dynamic structure factor [S28], namely

$$\int_{-\infty}^{\infty} D_{\pm\mp}(\omega)d\omega = \langle S_\pm S_\mp \rangle \equiv \left\langle \boldsymbol{S}_\perp^2 \right\rangle \pm \langle S_\| \rangle,$$
$$\int_{-\infty}^{\infty} \left(1 - e^{-\beta\hbar\omega}\right) D_{\pm\mp}(\omega)d\omega = \langle [S_\pm, S_\mp] \rangle \equiv \pm 2\langle S_\| \rangle,$$
$$(S64)$$

These properties can easily be obtained from the definitions of $D_{\pm\mp}(\omega)$ in Eq. (S46) and with the help of the detailed-balance relation in Eq. (S47). We emphasize that these properties are unrelated to RPA, and thus, the expression in Eq. (S63) is totally general in the elastic-collision limit. One can use any method to compute the static average $\langle \boldsymbol{S}_\perp^2 \rangle$. It is important to note that, in the special case of $S = 1/2$, one has $\langle \boldsymbol{S}_\perp^2 \rangle \equiv 1/2$ and the spin-flip rates are independent of the state of the magnetic system, in the elastic-collision approximation. In this case, no magnetic-field dependence of the spin-flip rates occurs. In general, for an arbitrary spin $S > 1/2$, the quantity $\langle \boldsymbol{S}_\perp^2 \rangle$ does depend on the magnetic field and assumes a maximum and a minimum in the following two extreme cases

$$\left\langle \boldsymbol{S}_\perp^2 \right\rangle = \begin{cases} \frac{2}{3}S(S+1), & \left\langle S_\| \right\rangle \to 0, \\ S, & \left\langle S_\| \right\rangle \to S. \end{cases} \quad (S65)$$

As a result, the spin-flip rates decrease for $S > 1/2$ with polarizing the magnetic system. However, the suppression does not occur here due to the energy conservation constraint. In the elastic-collision approximation, the electron has always sufficient energy to flip the spin. The reduction of the spin-flip rate is related to the squeezing of the state of $\boldsymbol{S}$, occurring when the magnetic system becomes polarized (e.g. by applying a magnetic field). The limit $\left\langle S_\| \right\rangle \to S$ corresponds to the most classical-like spin state possible, which is called the Bloch coherent state. In this state, the spin $\boldsymbol{S}$ has the least possible characteristic size of its transverse components. In addition to this "squeezing effect", the spin-flip rates are also affected beyond Eq. (S63) by the energy conservation law at low temperatures. This is clearly seen in the RPA result in Eq. (S57), where the rate contains the factor $n_B(\omega_q)$, which leads to a strong suppression of the rate with lowering the temperature, especially in a magnetic field due to the magnon gap.

In the elastic-collision approximation, the longitudinal spin relaxation rate reads

$$\frac{1}{\tau_\|} = \frac{1}{\tau_s} + \frac{\pi\nu J_{sd}^2}{\hbar b} n_{imp}^{2D} \left\langle \boldsymbol{S}_\perp^2 \right\rangle, \quad (S66)$$

where we added $1/\tau_s$ to account for the contribution of other spin-relaxation mechanisms in the metal, unrelated to scattering on local moments, see also the first term in Eq. (S2). As we shall see later, the term $1/\tau_s$ in Eq. (S66) drops out when we take the limit $b \to 0$ to obtain the boundary conditions.

Next, we turn to the transverse spin relaxation time $\tau_\perp$, i.e. the relaxation time of the transverse spin components of the itinerant electrons. The derivation goes along the same lines as above, except that the initial and final states in the Fermi golden rule are not the eigenstates of $\boldsymbol{n}\cdot\boldsymbol{\sigma}$, but superpositions of those states. Indeed, we are interested in the decay of an itinerant spin accumulation created along a direction in space, which is not collinear with $\boldsymbol{n}$. At the same time, the magnetic system remains polarized along $\boldsymbol{n}$ as before. It is not difficult to see that the spin decay will not be governed only by the ladder operators $S_\pm$, but also by the longitudinal spin operator $S_\|$, which appears now as "transverse" from the point of view of the itinerant spin accumulation. To be specific, we consider a fully transverse itinerant spin accumulation described by the spin states

$$|\Uparrow\rangle = \frac{1}{\sqrt{2}} \left( e^{-i\varphi/2} |\uparrow\rangle + e^{i\varphi/2} |\downarrow\rangle \right),$$
$$|\Downarrow\rangle = \frac{1}{\sqrt{2}} \left( e^{-i\varphi/2} |\uparrow\rangle - e^{i\varphi/2} |\downarrow\rangle \right), \quad (S67)$$

where $|\uparrow\rangle$ and $|\downarrow\rangle$ are the states in Eq. (S23) and $\varphi = \omega_L t + \varphi_0$ is the phase due to the Larmor precession, with $\varphi_0$ being an initial phase. We go to the rotating frame of the electron in which the Larmor precession is absent, but instead the local moments rotate relative to the electron spin in the opposite direction. Technically this is achieved by going to the interaction picture with the electron spin Hamiltonian $H = \frac{1}{2}\hbar\omega_L \boldsymbol{n}\cdot\boldsymbol{\sigma}$. The ladder operators $\frac{1}{2}\sigma_\pm$ acquire a time-dependent factor $e^{\pm i\omega_L t}$, whereas the states in Eq. (S67) stop evolving in time. Note that in this picture, the itinerant electrons lose their Zeeman and exchange splitting of the density of states. This mathematical procedure corresponds, loosely speaking, to moving the spin-dependent part of the exponent $e^{-\frac{it}{\hbar}(\varepsilon_{\boldsymbol{k}'\bar{s}} - \varepsilon_{\boldsymbol{k}s})}$ in Eq. (S18) from the electron transition energy to the time dependence of $U_{\text{int}}(t)$. Moreover, in the time-dependent $U_{\text{int}}(t)$ one swaps the exponents $e^{\pm i\omega_L t}$ from the itinerant spin ladder operators $\frac{1}{2}\sigma_\pm$ to the local moment ladder operators $S_\pm$. Thus, $S_\pm$ acquire an additional time dependence, which is not due to the magnetic system, but due to that, fact that relative to the electron, the local moments precess in the opposite direction,

$$S_\pm(t) = e^{\mp i\omega_L t} S_\pm. \tag{S68}$$

Mathematically this is evident from the relation

$$\boldsymbol{\sigma} \cdot \boldsymbol{S} = \frac{1}{2}\left(\sigma_+ S_- + \sigma_- S_+\right) + \sigma_\| S_\|, \tag{S69}$$

which shows that, *e.g.*, the time dependence from $\sigma_+$ is pulled over to $S_-$, which explains how the "opposite direction" comes about mathematically. Note also that $S_\|$ does not acquire any additional time dependence, since $\sigma_\|$ is static.

Following along the same lines of derivation as for the longitudinal spin relaxation time, we arrive at the analog of Eq. (S45), but now for the two spin-flip rates in the rotating fame ($w_{\Downarrow\Uparrow} = w_{\Uparrow\Downarrow}$)

$$w_{\Downarrow\Uparrow} = \frac{\pi n_{imp}^{3D} \nu J_{sd}^2}{2\hbar T} \iint_{-\infty}^{\infty} d\omega dE f(E)\left[1 - f(E - \hbar\omega)\right]$$
$$\times \left\{ D_{\|\|}(\omega) + \frac{1}{2}\left[D_{-+}(\omega) + D_{+-}(\omega)\right]\right\}, \tag{S70}$$

where

$$D_{\|\|}(\omega) = \frac{1}{2\pi}\int_{-\infty}^{\infty} dt e^{i\omega t} \left\langle S_\|(t) S_\|\right\rangle, \tag{S71}$$

and we have assumed axial symmetry in the magnetic system. The axial symmetry forbids the occurrence of certain correlators, such as between $S_\|$ and $S_\pm$, for instance, and makes the spin-flip rates independent of the starting phase $\varphi_0$ in Eq. (S67). The main difference in Eq. (S71) with respect to Eq. (S45) is the presence of the dynamic structure factor $D_{\|\|}(\omega)$. Unlike in the case of $D_{\pm\mp}(\omega)$, the longitudinal spin component $S_\|$ may have a finite average value, which is the order parameter $\langle S_\|\rangle$ of the magnetic system. This leads to an elastic-scattering channel in the rate in Eq. (S70). Indeed, we see this if we decompose $D_{\|\|}(\omega)$ following the general prescription [S28]

$$D_{\|\|}(\omega) = \left|\langle S_\|\rangle\right|^2 \delta(\omega) + \delta D_{\|\|}(\omega), \tag{S72}$$

where

$$\delta D_{\|\|}(\omega) = \frac{1}{2\pi}\int_{-\infty}^{\infty} dt e^{i\omega t}\left\langle \delta S_\|(t)\delta S_\|\right\rangle. \tag{S73}$$

Here, we denote $\delta S_\| = S_\| - \langle S_\|\rangle$, which is the operator of longitudinal spin fluctuations. Note that the first term on the right-hand size in Eq. (S72) is proportional to $\delta(\omega)$ and thus describes purely elastic scattering events.

Despite the fact that the division in Eq. (S72) is rather formal, it is possible to interpret the elastic-scattering terms as follows. Consider an electron with the spin oriented transversely to $\boldsymbol{n}$, *e.g.* a state such as in Eq. (S67). When moving randomly through the sample, at every scattering event, this electron picks a precession phase

$$\delta\varphi = \frac{E_{ex}\tau_c}{\hbar} \tag{S74}$$

by interacting with the magnetic impurity during a time $\tau_c = \lambda_F/v_F$, as given by Eq. (S3), and with a characteristic strength of exchange energy $E_{ex} = -J_{sd}\langle S_\|\rangle/\lambda_F^3$ felt by the electron during the collision time. Indeed, the electron can be regarded as a wave packet of size $\lambda_F$ along the motion of the electron and of area $\lambda_F^2$ in the transverse direction. In such a semiclassical picture, the electron sweeps along its trajectory a tube with cross section $\lambda_F^2$, as depicted in

$$\Delta\omega_L^{ex} = \frac{2\pi}{T} = \frac{\delta\varphi}{\tau_M} = -\frac{1}{\hbar}J_{sd}n_{imp}^{3D}\langle S_\parallel\rangle, \tag{S75}$$

which is also called sometimes the exchange correction to the Larmor frequency, since it occurs due to the $sd$-exchange coupling. It is remarkable that the result in Eq. (S75) doe not depend on the assumptions we made about the size of the wave packet. It can also be derived independently by averaging over the impurities positions $\boldsymbol{r}_j$ in Eq. (S5) and over the state of the magnetic system. On the other hand, the variance of the precession phase can be used to estimate the dephasing time. Unlike the case of Eq. (S75), the result is not universal, since it depends on the assumption made about the wave packet. We set $\sqrt{\langle(\varphi - \langle\varphi\rangle)^2\rangle} \sim 1$ to find the dephasing time: the characteristic time to lose track of the precession phase. Letting $T = \tau_\phi$ and hence $N = \tau_\phi/\tau_M$, we obtain

$$\frac{1}{\tau_\phi} \sim \frac{\delta\varphi^2}{\tau_M}. \tag{S76}$$

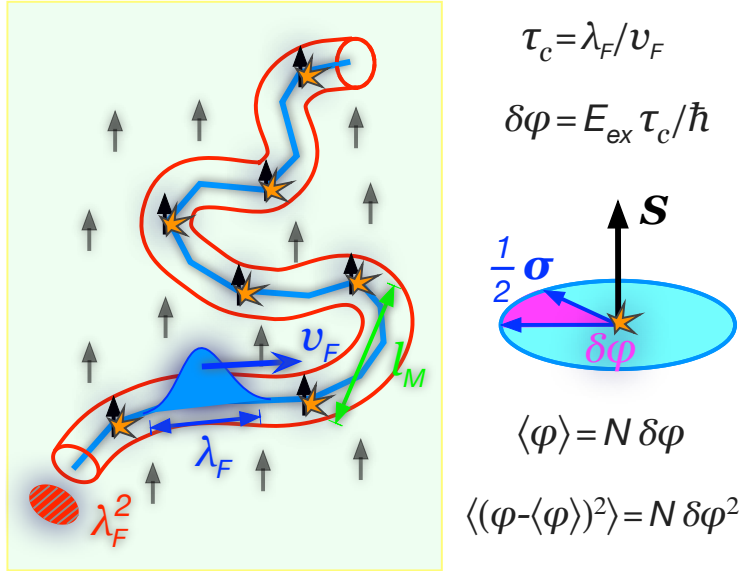

FIG. S10. Semiclassical picture of electron scattering off magnetic impurities. The electron is a wave packet of size $\lambda_F$ moving with speed $v_F$ and sweeping along its trajectory a tube of cross-section $\lambda_F^2$. The average distance between magnetic impurities along the tube is $l_M = 1/n_{imp}^{3D}\lambda_F^2$ and the average time between collisions is $\tau_M = l_M/v_F$. During the collision, the spin of the electron undergoes a short, but intense spike of Larmor precession, resulting in its precession phase being changed by $\delta\varphi$, each time in the same direction. After a time $T$, the precession phase acquires an average value $\langle\varphi\rangle = N\delta\varphi$ and a variance $\langle(\varphi - \langle\varphi\rangle)^2\rangle = N\delta\varphi^2$, where $N = T/\tau_M$.

Fig. S10. The magnetic impurities occur on the way of the electron randomly, at an average interval of time $\tau_M = 1/n_{imp}^{3D}\lambda_F^2 v_F$. The number of impurities met by the electron in during a given time $T$ is distributed according to the Poisson statistics. Thus, the average precession phase accumulated by the electron over time is $\langle\varphi\rangle = N\delta\varphi$, whereas the variance of this phase is $\langle(\varphi - \langle\varphi\rangle)^2\rangle = N\delta\varphi^2$, where $N = T/\tau_M$. Equating $\langle\varphi\rangle = 2\pi$, we obtain the average Larmor precession frequency due interaction with magnetic impurities

Up to numerical coefficients order unity, this simple exercise [S33] recovers the expresion for the dephasing time as obtained from the spin-flip rates in Eq. (S70) when keeping only the purely elastic-scattering channel of Eq. (S72). This agreement suggests that the elastic-scattering channel in Eq. (S72) describes spin dephasing due to the longitudinal exchange picked by the electron at impurities. The origin of this dephasing is in the Poisson statistics of the electron traveling times between magnetic impurities.

It is instructive to represent the elastic-channel term in a form in which the validity of the Born-Markov approximation is visible. After multiplying Eq. (S76) by a missing factor of $2\pi$, we have

$$\left(\frac{1}{\tau_\phi}\right)_{\text{el.-ch.}} = 2\pi \frac{\delta\varphi^2}{\tau_M} = \lambda_F^3 n_{imp}^{3D} \frac{E_{ex}^2}{\hbar/\tau_c} \frac{2\pi}{\hbar}. \tag{S77}$$

The factor $\lambda_F^3 n_{imp}^{3D}$ on the right-hand side here is the probability that the electron (*i.e.* a wave packet of volume $\lambda_F^3$) has encountered a magnetic impurity at a given moment in time. This probability is present in Eq. (S77) because we are adding incoherently the contribution of different impurities, assuming that they are sufficiently far away from each other. One could incorporate this factor as a square root into $E_{ex}$ to make for an effective interaction strength of a collection of incoherent sources. The factor $E_{ex}^2 \tau_c/\hbar$ in Eq. (S77) describes what happens when the electron has encountered a magnetic impurity. The Born-Markov approximation is clearly visible here: one squares the interaction energy $E_{ex}$ and divides it by the characteristic energy of the bath $\hbar/\tau_c$, where $\tau_c$ has the meaning of the bath correlation time. Clearly, the local moments are a relatively slow system and would not qualify on their own for a Markovian bath. The electron motion introduces the short time scale $\tau_c$, which makes the Born-Markov approximation valid. This reduction is commonly termed *motional narrowing*. We conclude this pedagogical remark by noting that the Born approximation requires that $\lambda_F^3 n_{imp}^{3D} \ll 1$, which is the condition for diluteness of magnetic impurities, whereas the Markovian limit means $E_{ex} \ll \hbar/\tau_c$, which is equivalent to $\nu J_{sd} \ll 1$ and is the condition for weak coupling between a magnetic impurity and an itinerant electron at the Fermi level. Despite the fact that the diluteness condition seemingly appears to not be satisfied when taking $b \to 0$, it is obvious that the mathematical $b \to 0$ means physically $b \to \lambda_F$, since the Fermi energy is the highest ultraviolet cutoff we attend to in the metal.

As for the second term on the right-hand side in Eq. (S72), RPA does not permit to compute it right away [S20, S26]. One can use the results of the diagramatic technique for spins [S30], in which the correlator of longitudinal spin fluctuations can be approximated by a certain series of diagrams. Far away from the critical point, one obtains the following physical picture for excitation of collective modes. Clearly, since the longitudinal spin fluctuations converse the total longitudinal spin of the magnetic system, no magnons can be created or destroyed by means of coupling to the operator $S_\parallel$. However, energy can be given (or taken) from the magnetic system by a process in which an already existing magnon in the system is promoted up (or demoted down) in energy. Since at low temperatures most magnons have an almost vanishing energy, one can assume that the magnons are promoted up in energy from $\omega_q \approx 0$. The absorption spectrum will therefore have a somewhat smeared peak at an energy branch which coincides with the magnon excitation spectrum [S30]. And vice versa, since most magnons are at low energies, the emission spectrum will be peaked near $\omega \approx 0$, on the side $\omega < 0$. Introducing as in Eq. (S50) the retarded linear response function

$$\chi_\parallel(t) = -\frac{i}{\hbar}\theta(t)\left\langle\left[\delta S_\parallel(t), \delta S_\parallel\right]\right\rangle, \tag{S78}$$

we use the results of Ref. S30 to write

$$\chi_{\parallel\parallel}(\omega) = \frac{1}{M^2\hbar}\sum_{qq'} \frac{n_B(\omega_{q'}) - n_B(\omega_q)}{\omega - \omega_q + \omega_{q'} + i0}, \tag{S79}$$

which we use further in a similar way as we used Eq. (S52). Namely, we write down the fluctuation-dissipation theorem [S28],

$$\delta D_{\|\|}(\omega) = -\frac{\hbar}{\pi}\left[1 + n_B(\omega)\right]\Im\left[\chi_{\|\|}(\omega)\right], \tag{S80}$$

to obtain

$$\delta D_{\|\|}(\omega) = \frac{1}{M^2}\sum_{\boldsymbol{q}\boldsymbol{q}'}\left[1 + n_B(\omega_{\boldsymbol{q}} - \omega_{\boldsymbol{q}'})\right] \\ \times \left[n_B(\omega_{\boldsymbol{q}'}) - n_B(\omega_{\boldsymbol{q}})\right]\delta(\omega - \omega_{\boldsymbol{q}} + \omega_{\boldsymbol{q}'}). \tag{S81}$$

For $\omega > 0$, this expression describes the absorption of a quanta of energy $\hbar\omega$ by the magnetic system by means of promoting (thermal) magnons up in energy from states $\boldsymbol{q}'$ with a lower energy to states $\boldsymbol{q}$ with a higher energy. The absorption rate is proportional to $1+n_B(\omega)$, which shows that the magnetic system can always absorb energy (within the range of the magnon bandwidth), provided thermal magnons exist in the system to begin with. Conversely, for $\omega < 0$, the quantity $\delta D_{\|\|}(\omega)$ describes the emission of a quanta of energy $\hbar\omega$ by the magnetic system by means of demoting (thermal) magnons down in energy from states $\boldsymbol{q}'$ with a higher energy to states $\boldsymbol{q}$ with a lower energy. The emission rate is proportional to $n_B(\omega)$, which shows that the magnetic system can only give away energy if the temperature is not zero, *i.e.* the system is not in its ground state.

It is important to remark that Eq. (S81) is derived outside of RPA and does not, therefore, comply with the spectral sum rule. From Eq. (S81), we obtain

$$\int_{-\infty}^{\infty}\delta D_{\|\|}(\omega)d\omega = \frac{1}{M}\sum_{\boldsymbol{q}}n_B(\omega_{\boldsymbol{q}}) \equiv m. \tag{S82}$$

On the other hand, by the definition of $\delta D_{\|\|}(\omega)d\omega$ in Eq. (S73), we have the spectral sum rule

$$\int_{-\infty}^{\infty}\delta D_{\|\|}(\omega)d\omega = \langle \delta S_{\|}^2 \rangle = \langle S_{\|}^2 \rangle - \langle S_{\|} \rangle^2. \tag{S83}$$

The averages $\langle S_{\|}^2 \rangle$ and $\langle S_{\|} \rangle$ can be expressed using the RPA approximation in terms of $m$ and $S$. The expression for the right-hand side of Eq. (S83) is a rather cumbersome function of $m$ and $S$. We expand it for small $m$, obtaining

$$\langle S_{\|}^2 \rangle - \langle S_{\|} \rangle^2 = \begin{cases} m - 3m^2 + \mathcal{O}(m^3), & S = 1/2, \\ m + m^2 + \mathcal{O}(m^{2S+1}), & S \geq 1. \end{cases} \tag{S84}$$

As we see, in the limit $m \ll 1$, the spectral weight in Eq. (S82) agrees with the one expected from the spectral sum rule in Eq. (S83) and with the static averages calculated using RPA. For more elaborate approximations to the longitudinal spin propagator, we refer the reader to Ref. S34.

Using the rate in Eq. (S70), we obtain for the transverse spin relaxation time the well-known formula [S23]

$$\frac{1}{\tau_\perp} = \frac{1}{2\tau_\|} + \frac{1}{\tau_\phi}, \tag{S85}$$

where $1/\tau_\|$ is the longitudinal spin relaxation rate calculated above and $1/\tau_\phi$ is the rate termed as *secular broadening*. It is given by

$$\frac{1}{\tau_\phi} = \frac{\pi n_{imp}^{3D} \nu J_{sd}^2}{\hbar} \left[ \langle S_\| \rangle^2 + \frac{1}{T} \iint_{-\infty}^{\infty} d\omega dE f(E) \left[1 - f(E - \hbar\omega)\right] \delta D_{\|\|}(\omega) \right]. \tag{S86}$$

Using the expression in Eq. (S81), we obtain

$$\frac{1}{\tau_\phi} = \frac{\pi n_{imp}^{3D} \nu J_{sd}^2}{\hbar} \left[ \langle S_\| \rangle^2 + \frac{2\hbar}{TM^2} \sum_{\boldsymbol{qq'}} (\omega_{\boldsymbol{q}} - \omega_{\boldsymbol{q'}}) n_B(\omega_{\boldsymbol{q}} - \omega_{\boldsymbol{q'}}) \left[1 + n_B(\omega_{\boldsymbol{q}} - \omega_{\boldsymbol{q'}})\right] n_B(\omega_{\boldsymbol{q'}}) \right], \tag{S87}$$

which can also be written more compactly as

$$\frac{1}{\tau_\phi} = \frac{\pi n_{imp}^{3D} \nu J_{sd}^2}{\hbar} \left[ \langle S_\| \rangle^2 - \frac{2\beta}{M^2} \sum_{\boldsymbol{qq'}} n_B(\omega_{\boldsymbol{q'}}) \frac{\partial}{\partial \beta} n_B(\omega_{\boldsymbol{q}} - \omega_{\boldsymbol{q'}}) \right]. \tag{S88}$$

One can view Eq. (S88) as a perturbative expansion in multi-magnon processes with the second term in the square brackets being the contribution of the two-magnon process. In Appendix A, we identify three regimes of applicability of Eq. (S88) for the 2D Heisenberg ferromagnetic model and provide expressions for the asymptotic expansion of Eq. (S88) in these regimes. Our analysis shows that the dephasing rate is dominated by the elastic-scattering channel at low temperatures ($T \ll SJ$), provided the magnetic field is sufficiently high to saturate the magnetization. We emphasize that none of these conditions are generally met in the experiment and Eq. (S88) cannot be used to fit the experimental data.

We turn now to the high-temperature limit, in which the itinerant electrons can exchange with the magnetic system any energy within the magnon bandwidth ($\sim \langle S_\| \rangle J$) and the transition not be blocked by the Pauli exclusion principle. We proceed along the same lines as around Eq. (S62) and obtain

$$\frac{1}{\tau_\phi} = \frac{\pi n_{imp}^{3D} \nu J_{sd}^2}{\hbar} \langle S_\|^2 \rangle. \tag{S89}$$

Here, we used the spectral sum rule in Eq. (S83) and a similar property to the second line of Eq. (S64),

$$\int_{-\infty}^{\infty} \left(1 - e^{-\beta\hbar\omega}\right) \delta D_{\|\|}(\omega) d\omega = 0. \tag{S90}$$

We emphasize that, for Heisenberg models in low dimensions, the high-temperature limit does not necessarily reduce to the paramagnetic limit. The condition of a small magnon bandwidth can be satisfied at arbitrarily low temperatures by lowering the magnetic field sufficiently, such that $\langle S_\| \rangle J \ll T$. On the other hand, the true paramagnetic limit is achieved in these models for $T \gg T_W$, where $T_W \sim S^2 J$ is the energy scale which plays the role of the Curie temperature in the Weiss mean-field theory. We can, therefore, use RPA to calculate the static average $\langle S_\|^2 \rangle$ in Eq. (S89) and reproduce, as a result, effects due to the ferromagnetic correlations developing in the low-dimensional magnetic system at the experimentally relevant temperatures (which happen to be $T \lesssim T_W$).

To summarize, we obtained for the longitudinal and transverse spin-relaxation times in the high-temperature regime, $T \gg \langle S_\| \rangle J, g\mu_B B$, the following expressions

$$\frac{1}{\tau_\|} = \frac{1}{\tau_s} + \frac{\pi \nu J_{sd}^2}{\hbar b} n_{imp}^{2D} \langle \boldsymbol{S}_\perp^2 \rangle,$$

$$\frac{1}{\tau_\perp} = \frac{1}{\tau_s} + \frac{\pi \nu J_{sd}^2}{\hbar b} n_{imp}^{2D} \left( \langle S_\|^2 \rangle + \frac{\langle \boldsymbol{S}_\perp^2 \rangle}{2} \right). \tag{S91}$$

It is convenient to use the Casimir invariant to relate $\langle \bm{S}_\perp^2 \rangle$ to $\langle S_\parallel^2 \rangle$,

$$S(S+1) = \bm{S}_\perp^2 \rangle + \langle S_\parallel^2 \rangle. \tag{S92}$$

To calculate $\langle S_\parallel^2 \rangle$, we use RPA as explained above.

Besides the appearance of the finite spin relaxation tensor due to the magnetic interface, one also has to take into account the corrections to the Larmor precession frequency $\omega_L$ due to the magnetic polarization in the interface region. For a magnetization collinear with the magnetic field, the renormalized Larmor frequency reads [c.f. Eq. (S75)]

$$\hbar\omega_L(z) = g\mu_B B - J_{sd} n_{imp}^{2D} \langle S_\parallel \rangle \delta_b(z), \tag{S93}$$

where $\delta_b(z)$ equals to $1/b$ in the $b$-region and zero elsewhere. In the limit $b \to 0$, $\delta_b(z)$ tends to the $\delta$-function and describes the effect of the interface exchange interaction induced by the ferromagnetic insulator on the itinerant electrons in the metal. The interface exchange interaction is known to be important in certain metal/ferromagnetic insulator sandwiches, see, e.g., Ref. S36 for a recent study of the Al/EuS system. Note that the correction to $\omega_L$ in the $b$-region is the leading order effect coming from the magnetic interface since it is linear in the $s$-$d$ coupling $J_{sd}$. In contrast, the rates $1/\tau_\parallel$ and $1/\tau_\perp$ contain magnetic terms only at the second-order in $J_{sd}$.

Now we integrate Eq. (S1) over the boundary layer $b$. by taking into account Eqs. (S4) and (S93) and obtain

$$\frac{1}{e\nu} J_{zj} \Big|_{z=0}^{z=b} = \frac{b}{\tau_\perp} \mu_{s,j} + \left(\frac{b}{\tau_\parallel} - \frac{b}{\tau_\perp}\right) n_j (\bm{n}\cdot\bm{\mu}_s) \\ - b\Omega_L \varepsilon_{jik} n_i \mu_{s,k}, \tag{S94}$$

where

$$\hbar\Omega_L = g\mu_B B - \frac{J_{sd} n_{imp}^{2D}}{b} \langle S_\parallel \rangle. \tag{S95}$$

By requiring that the spin current vanishes at the outer interface $z = 0$, we obtain the boundary condition for the spin current at $z = b$, which in vector notation reads

$$\frac{1}{e\nu} \bm{J}_z = \frac{b}{\tau_\parallel} \bm{\mu}_s + \left(\frac{b}{\tau_\parallel} - \frac{b}{\tau_\perp}\right) \bm{n} \times [\bm{n} \times \bm{\mu}_s] \\ - b\Omega_L \bm{n} \times \bm{\mu}_s. \tag{S96}$$

This boundary condition can be written in a more customary way

$$-e\bm{J}_z = G_s \bm{\mu}_s + G_r \bm{n} \times [\bm{n} \times \bm{\mu}_s] + G_i \bm{n} \times \bm{\mu}_s, \tag{S97}$$

where we have introduced the spin-mixing conductances:

$$G_s = -\frac{\nu b e^2}{\tau_\parallel},$$
$$G_r = \nu b e^2 \left(\frac{1}{\tau_\perp} - \frac{1}{\tau_\parallel}\right),$$
$$G_i = \nu b e^2 \Omega_L. \tag{S98}$$

Taking the limit $b \to 0$, we see that the term $1/\tau_s$ in Eq. (S91) does not contribute to the boundary condition. We obtain

$$G_s = -\frac{\pi e^2 \nu^2 J_{sd}^2}{\hbar} n_{imp}^{2D} \langle \boldsymbol{S}_\perp^2 \rangle,$$

$$G_r = \frac{\pi e^2 \nu^2 J_{sd}^2}{\hbar} n_{imp}^{2D} \left( \langle S_\parallel^2 \rangle - \frac{\langle \boldsymbol{S}_\perp^2 \rangle}{2} \right),$$

$$G_i = -\frac{e^2 \nu J_{sd}}{\hbar} n_{imp}^{2D} \langle S_\parallel \rangle. \tag{S99}$$

We recall that $S_\parallel$ is the component of spin in the direction of the magnetic field and $\boldsymbol{S}_\perp$ are the transverse components, see Eq. (S11). If the magnetic filed is directed along the $z$-axis, then we have $S_\parallel = S_z$ and $\boldsymbol{S}_\perp = (S_x, S_y, 0)$.

The boundary condition Eq. (S97) coincides with the one derived in Ref. [S37] and used for the study of the spin Hall magnetoresistance (SMR) effect in magnetic-insulator/metal bilayers [S38]. In that latter work, however, the dependence of the spin conductances on temperature and on the magnetic field have not been taken into account. The novelty in our model is that we have expressed the $G$'s in Eq. (S97) in terms of the spin operator averages, Eq. (S99). We can then obtain not only the $T$ and $B$ dependence of the spin-mixing conductance, but also describe different magnetic materials and textures. The information about the latter is encoded in the spin operator averages.

The boundary condition (S97) together with the expressions (S99) have been used in the main text to calculate the magnetoresistance.

**Appendix A: Three regimes of applicability of Eq. (S88)**

We consider here the second term in the square brackets in Eq. (S88), denoting it as

$$\text{VLP} := -\frac{2\beta}{M^2} \sum_{\boldsymbol{q}\boldsymbol{q}'} n_B(\omega_{\boldsymbol{q}'}) \frac{\partial}{\partial \beta} n_B(\omega_{\boldsymbol{q}} - \omega_{\boldsymbol{q}'}). \tag{A1}$$

The applicability of Eq. (S88) is bound to the condition $m \ll 1$, which physically means that there are few magnons in the system. The integrals arising from the double sum over the magnon wave vectors in Eq. (A1) gain their weight near the lower magnon band edge. Considering a square lattice with nearest-neighbor interaction, we expand the magnon spectrum for small $\boldsymbol{q} = (q_x, q_y)$,

$$\hbar \omega_{\boldsymbol{q}} \approx J \left| \langle S_\parallel \rangle \right| (q_x^2 + q_y^2) + g\mu_B B, \tag{A2}$$

and obtain after integration

$$\text{VLP} \approx -\frac{(1+\gamma_E)\ln\left(1 - e^{-\beta g \mu_B B}\right) + F\left(e^{-\beta g \mu_B B}\right)}{8\pi^2 \beta^2 J^2 \langle S_\parallel \rangle^2}, \tag{A3}$$

where $\gamma_E \approx 0.57$ is Euler's constant and $F(x)$ is a function, which can be expressed in terms of the polylogarithm function $\text{Li}_n(x)$ differentiated over its order $n$,

$$F(x) := \lim_{n \to 1} \frac{\partial}{\partial n} \text{Li}_n(x) = -\sum_{k=1}^{\infty} \frac{x^k}{k} \ln k. \tag{A4}$$

The polylogarithm derivative dominates in Eq. (A3) for small Zeeman energies ($g\mu_B B \ll T$), whereas the logarithm dominates for large Zeeman energies ($g\mu_B B \gg T$). In the latter case, the two-magnon process is exponentially suppressed

$$\text{VLP} \approx \frac{1+\gamma_E}{8\pi^2}\left(\frac{T}{SJ}\right)^2 \exp\left(-\frac{g\mu_B B}{T}\right). \quad (A5)$$

This result is valid for $T \ll SJ$, since, at these large magnetic fields, the magnon band attains its full width, with $\langle S_\parallel \rangle = -S$, and the magnons can be promoted up in energy only close to the lower band edge. For intermediate temperatures, $SJ \ll T \ll g\mu_B B$, the approximation in Eq. (A2) is invalid and so is the result in Eq. (A3). We go back to Eq. (A1) and compute it by expanding $n_B(\omega_{\bm q} - \omega_{\bm q'})$ for large temperatures, obtaining

$$\text{VLP} \approx \exp\left(-\frac{g\mu_B B}{T}\right). \quad (A6)$$

Clearly, Eqs. (A5) and (A6) show that, for large Zeeman energies, the two-magnon process represents a small correction on top of the elastic-scattering channel.

In the opposite limiting case, for small Zeeman energies ($g\mu_B B \ll T$), we keep only the leading-order term in the asymptotic expansion of the polylogarithm derivative, $F(e^{-x}) \approx -\frac{1}{2}\ln^2 x$, for $x \ll 1$. It is important to realize that the temperature itself must be sufficiently small in order for the regime $g\mu_B B \ll T$ to be compatible with the condition $m \ll 1$. With the quadratic approximation in Eq. (A2), we obtain

$$m \approx \frac{T}{4\pi J \langle S_\parallel \rangle} \ln \frac{g\mu_B B}{T}. \quad (A7)$$

Then, we approximate $\langle S_\parallel \rangle \approx -\chi B/g\mu_B$, where $\chi$ is the susceptibility which was introduced in Eq. (S60). Assuming that the temperature is sufficiently low, such that $\chi$ is sufficiently high, we obtain that the condition $m \ll 1$ translates to

$$g\mu_B B \gg T\frac{(g\mu_B)^2}{4\pi J \chi} \ln \frac{4\pi J \chi}{(g\mu_B)^2}. \quad (A8)$$

For $\chi$, we use the expression derived in Ref. S26 in the limit of small temperatures ($T \ll T_W$)

$$\chi \simeq \frac{(g\mu_B)^2}{32 J} e^{\pi \frac{T_W}{T}}, \quad (A9)$$

where $T_W = \frac{4}{3}S(S+1)J$ is the critical temperature appearing in the Weiss theory [S35]. We see that for sufficiently low $T$, the scale in Eq. (A8) can be made much smaller than the temperature, so that a regime of $B$-fields emerges where $m \ll 1$, on the one hand, and $g\mu_B B \ll T$, on the other hand. For this to happen, it is sufficient to have $T \ll SJ$, where we recalled that we used the quadratic approximation for the magnon dispersion and that $SJ < T_W$. The scale in Eq. (A8) can then be written as

$$g\mu_B B \gg T_W e^{-\pi \frac{T_W}{T}}. \quad (A10)$$

One can verify that the magnetic field is sufficiently high in this regime in order to almost fully polarize the spin. Hence, we approximate $\langle S_\parallel \rangle \approx -S$. As a result, we obtain for the two-magnon process

$$\text{VLP} \approx \frac{1}{16\pi^2}\left(\frac{T}{SJ}\right)^2 \ln^2 \frac{T}{g\mu_B B}, \quad (A11)$$

which is valid in the regime

$$SJ e^{-\pi \frac{T_W}{T}} \ll g\mu_B B \ll T \ll SJ. \quad (A12)$$

Unlike in the case of large Zeeman energies, the regime $T \gg SJ$ cannot be considered here, since it invalidates our use of the two-magnon process. Indeed, the condition $m \gg 1$ follows directly from $T \gg SJ, g\mu_B B$, see Eqs. (S13) and (S15). By comparing Eqs. (A5) and (A11) with each other, we see that, with decreasing the magnetic field, the exponential suppression of the two-magnon process present at large magnetic fields turns into a square-logarithmic enhancement at small magnetic fields.

To give a complete picture here, we need also to consider the first term in the square brackets in Eq. (S88). For $m \ll 1$, one obtains from Eq. (S16)

$$\langle S_\parallel \rangle^2 \approx S^2 - 2Sm + \begin{cases} -3m^2 + \mathcal{O}\left(m^3\right), & S = 1/2, \\ m^2 + \mathcal{O}\left(m^{2S+1}\right), & S \geq 1. \end{cases} \quad (A13)$$

Thus, we need to evaluate $m$ in the three regimes considered above. Using the quadratic approximation in Eq. (A2), we obtain [c.f. Eq. (A3)]

$$m \approx \frac{1}{4\pi\beta J \langle S_\parallel \rangle} \ln\left(1 - e^{-\beta g \mu_B B}\right). \quad (A14)$$

For large Zeeman energies ($g\mu_B B \gg T$), we have

$$m \approx \frac{1}{4\pi} \frac{T}{SJ} \exp\left(-\frac{g\mu_B B}{T}\right), \quad (A15)$$

which is valid for small temperatures $T \ll SJ$. At large temperatures and even larger Zeeman energies ($SJ \ll T \ll g\mu_B B$), we obtain directly from the definition of $m$ without using the quadratic approximation

$$m \approx \exp\left(-\frac{g\mu_B B}{T}\right). \quad (A16)$$

Finally, in the regime (A12), we have

$$m \approx \frac{1}{4\pi} \frac{T}{SJ} \ln \frac{T}{g\mu_B B}. \quad (A17)$$

Taking into account Eqs. (A13) and (S88), we conclude as follows.

In the first regime ($T \ll SJ, g\mu_B B$), the two-magnon process modifies weakly (in $T/SJ$) the prefactor of the exponential suppression with the Zeeman energy,

$$\frac{1}{\tau_\phi} \approx \frac{\pi n_{imp}^{3D} \nu J_{sd}^2}{\hbar} \left\{ S^2 - \left[\frac{T}{2\pi J} - \frac{1+\gamma_E}{8\pi^2}\left(\frac{T}{SJ}\right)^2\right] e^{-\frac{g\mu_B B}{T}} \right\}, \quad (A18)$$

where we highlighted the part of the prefactor coming from the two-magnon process. We see that its contribution constitutes here only a small portion ($\sim T/SJ$) of its full spectral weight [which is $m$ according to Eq. (S82)]. This reduction is explained by the low temperature ($T \ll SJ$) which makes the Fermi sea degenerate and, as a result, only electrons within a small energy range ($\sim T$) around the Fermi surface participate in inelastic scattering. Scattering events with large energy transfer cannot occur due to the Pauli exclusion principle. Hence, the electrons "probe" only a small portion of the spectral weight of $\delta D_{\parallel\parallel}(\omega)$ in a range of frequencies $\sim T/\hbar$ around $\omega = 0$.

In the second regime ($SJ \ll T \ll g\mu_B B$), the two-magnon process modifies the exponential prefactor stronger than in the previous regime,

$$\frac{1}{\tau_\phi} \approx \frac{\pi n_{imp}^{3D} \nu J_{sd}^2}{\hbar} \left[ S^2 - (2S-1) e^{-\frac{g\mu_B B}{T}} \right]. \quad (A19)$$

In particular, this has an important consequence for the case $S = 1/2$, for which the prefactor of the leading exponential dependence vanishes and one would expect a stronger suppression, at least as the exponent squared

$$\propto e^{-\frac{2g\mu_B B}{T}}. \quad (A20)$$

However, it appears that the prefactors of all such exponents vanish for $S = 1/2$. As a sanity check, we test this result in the paramagnetic limit, in which the correlator $\delta D_{\|\|}(\omega)$ is fully static

$$\delta D_{\|\|}(\omega) = \left( \langle S_\|^2 \rangle - \langle S_\| \rangle^2 \right) \delta(\omega), \quad (A21)$$

and we have

$$\frac{1}{\tau_\phi} = \frac{\pi n_{imp}^{3D} \nu J_{sd}^2}{\hbar} \langle S_\|^2 \rangle. \quad (A22)$$

Indeed, it follows from Eq. (A22) that for $S = 1/2$, there is no exponential dependence at all, since $S_\|^2 = 1/4$. Moreover, for higher spins, we recover the prefactor $(2S - 1)$ as in front of the exponent in Eq. (A19). To do this, we calculate $\langle S_\|^2 \rangle$ in the paramagnetic case and, after expanding for $g\mu_B B \gg T$, we obtain that Eq. (A22) yields the same asymptotic term as given in Eq. (A19). This is not surprising, since the condition $SJ \ll T$ corresponds essentially to the paramagnetic limit. Nonetheless, this agreement serves as evidence that the two-magnon process represents the leading-order correction to the elastic-scattering channel.

In the third regime, see Eq. (A12), the two-magnon process is of the same order as the terms $\propto m^2$ in Eq. (A13). Combing both terms together, we obtain

$$\frac{1}{\tau_\phi} \approx \frac{\pi n_{imp}^{3D} \nu J_{sd}^2}{\hbar} \left[ S^2 - \frac{T}{2\pi J} \ln \frac{T}{g\mu_B B} \right.$$
$$\left. \mp \frac{1}{8\pi^2} \left( \frac{T}{SJ} \right)^2 \ln^2 \frac{T}{g\mu_B B} \right], \quad (A23)$$

where the upper sign refers to $S = 1/2$ and the lower sign to $S \geq 1$. The last term arises from the combination of the elastic-scattering and two-magnon processes. Again, as in Eq. (A18), the two-magnon process contributes here with a suppression factor $\sim T/SJ$ due to the Fermi sea becoming degenerate at low temperatures. However, this suppression factor is partly compensated by a logarithmic enhancement factor $\ln(T/g\mu_B B)$ coming from the bosonic bunching factor $1 + n_B$ which plays a significant role here during the integration at the lower band edge.

**Supplemental References**

[S1] T. Shang, Q. F. Zhan, H. L. Yang, Z. H. Zuo, Y. L. Xie, Y. Zhang, L. P. Liu, B. M. Wang, Y. H. Wu, S. Zhang, and R.-W. Li, Phys. Rev. B **92**, 165114 (2015).

[S2] X. Jia, K. Liu, K. Xia, and G. E. W. Bauer, Europhys. Lett. **96**, 17005 (2011).

[S3] M. Isasa, A. Bedoya-Pinto, S. Vélez, F. Golmar, F. Sanchez, L. E. Hueso, J. Fontcuberta, and F. Casanova, Appl. Phys. Lett. **105**, 142402 (2014).

[S4] M. Isasa, S. Vélez, E. Sagasta, A. Bedoya-Pinto, N. Dix, F. Sánchez, L. E. Hueso, J. Fontcuberta, and F. Casanova, Phys. Rev. Appl. **6**, 034007 (2016).

[S5] H. Hsu, P. Blaha, and R. M. Wentzcovitch, Phys. Rev. B **85**, 140404(R) (2012).

[S6] M. S. D. Read, M. S. Islam, G. W. Watson, F. King, and F. E. Hancock., J. Mater. Chem. **10**, 2298 (2000).

[S7] S. Khan, R. J. Oldman, F. Corà, C. R. A. Catlow, S. A. French, and S. A. Axon, Phys. Chem. Chem. Phys. **8**, 5207 (2006).

[S8] Twagirashema, M. Frere, L. Gengembre, C. Dujardin, and P. Granger, Topics in Catalysis **42**, 171 (2007).

[S9] J. P. Dacquin, C. Dujardin, and P. Granger, Journal of Catalysis **253**, 37 (2008).

[S10] Z. Feng, Y. Yacoby, W. T. Hong, H. Zhou, M. D. Biegalski, H. M. Christen, and Y. Shao-Horn, Energy Environ. Sci. **7**, 1166 (2014).

[S11] X. Liu, Z. Chen, Y. Wen, R. Chen, and B. Shan, Catal. Sci. Technol. **4**, 3687 (2014).

[S12] Abdullah Radi, PhD thesis, *Soft X-ray Reflectometry and X-ray Absorption Spectroscopy of Transition Metal Oxide Thin-Films on Various Substrates: Cobaltate (CoO) and Lanthanum Cobaltate (LaCoO3)*, University of British Columbia, Vancouver (2017).

[S13] M. P. de Jong, I. Bergenti, V. A. Dediu, M. Fahlman, M. Marsi, and C. Taliani, Phys Rev B **71**, 014434 (2005).

[S14] J. M. Pruneda, V. Ferrari, R. Rurali, P. B. Littlewood, N. A. Spaldin, and E. Artacho, Phys. Rev. Lett. **99**, 226101 (2007).

[S15] C. A. Marianetti, G. Kotliar, and G. Ceder, Phys. Rev. Lett. **92**, 196405 (2004).

[S16] N. Biškup, J. Salafranca, V. Mehta, M. P. Oxley, Y. Suzuki, S. J. Pennycook, S. T. Pantelides, and M. Varela, Phys. Rev. Lett. **112**, 087202 (2014).

[S17] J.-Q. Yan, J.-S. Zhou, and J. B. Goodenough, Phys Rev B **70**, 014402 (2004).

[S18] L. Ronding, J.-P. Tetienne, T. Hingant, J.-F. Roch, P. Maletinsky, and V. Jacques, Rep. Prog. Phys. **77**, 056503 (2014).

[S19] S. Vélez, V. N. Golovach, A. Bedoya-Pinto, M. Isasa, E. Sagasta, M. Abadia, C. Rogero, L. E. Hueso, F. S. Bergeret, and F. Casanova, Phys. Rev. Lett. **116**, 016603 (2016).

[S20] N. Majlis, *The Quantum Theory of Magnetism* (World Scientific Publishing Co. Pte. Ltd., Singapore, 2007).

[S21] F. Rivadulla, Z. Bi, E. Bauer, B. Rivas-Murias, J. M. Vila-Fungueiriño, and Q. Jia, Chem. Mater. **25**, 55 (2013).

[S22] M. W. Wu, J. H. Jiang, and M. Q. Weng, Physics Reports **493**, 61 (2010).

[S23] C. P. Slichter, *Principles of Magnetic Resonance*, (Springer-Verlag, Berlin, 1980).

[S24] H. Callen, Phys. Rev. **130**, 890 (1963).

[S25] N. D. Mermin and H. Wagner, Phys. Rev. Lett. **17**, 1133 (1966).

[S26] D. A. Yablonskiy, Phys. Rev. B **44**, 4467 (1991).

[S27] One can show that $[S_+, S_-] = 2\mathbf{n}\cdot\mathbf{S}$ with the help of the usual commutation relation for the spin components in cartesian coordinates $[S_j, S_k] = i\epsilon_{jkl}S_l$ and noting that

$$\sum_{jk} \epsilon_{jkl} S_l \langle\downarrow|\sigma_j|\uparrow\rangle\langle\uparrow|\sigma_k|\downarrow\rangle = -2in_l.$$

[S28] Gabriele Giuliani and Giovanni Vignale, *Quantum Theory of the Electron Liquid*, (Cambridge University Press, 2005).

[S29] The magnon damping occurs in the diagramatic treatment of the problem [10, 11] at higher orders of $J_q$ and amounts to replacing the denominator $\omega - \omega_q + i0$ in Eq. (S52) by $\omega - \omega_q + i\gamma_q(\omega)$, where $\gamma_q$ is the magnon line-width, which is given to lowest order by

$$\gamma_q(\omega) = \frac{\pi}{M\hbar^2}\sum_{q'}\left[\frac{SB'_S(\beta h)\left(J_{q'-q}-J_{q'}\right)^2}{1-\beta SB'_S(\beta h)J_{q'-q}}\delta(\omega-\omega_{q'})\right],$$

with $h = g\mu_B B - \langle S_{||}\rangle J_0$ being the Weiss field and $B'_S(x)$ the derivative of the Brillouin function.

[S30] V. G. Vaks, A. I. Larkin, and S. A. Pikin, Zh. Eksp. Teor. Phys. **53**, 1089 (1967) [Sov. Phys. JETP **26**, 647 (1968)].

[S31] Yu. A. Izyumov and Yu. N. Skryabin, *Statistical Mechanics of Magnetically ordered systems*, (Consultants Bureau New York and London, 1988).

[S32] The expansion in Eq. (S62) is based on the identity

$$\frac{t}{1+t+\frac{p}{1-p}} = \frac{t}{1+t}\left[1 - \frac{p}{1+t}\left(\frac{1}{1-p\frac{t}{t+1}}\right)\right].$$

Expanding here the factor in parentheses into a geometric series one obtains the required expansion in Eq. (S62). After identifying $t \equiv e^{\beta(E-\hbar)}$ and $p \equiv 1 - e^{-\beta\hbar\omega}$, it becomes clear that the left-hand side of this identity is $1 - f(E-\hbar\omega)$.

[S33] Using the following expression for the density of states $\nu = 2/\hbar\lambda_F^2 v_F$, one recovers the result of the golden-rule calculation up to a factor of $2\pi$, which could be absorbed into a rededefinition of the cross section of the tube in Fig. S10. An area of $\lambda_F^2/2\pi$ turns out to be a perfect choice.

[S34] Yu. A. Izyumov, N. I. Chaschin, and V. Yu. Yushankhai, Phys. Rev. B **65**, 214425 (2002).

[S35] In RPA applied to the two-dimensional Heisenberg ferromagnet, TW has the meaning of a crossover temperature scale between the weakly and strongly correlated regimes. In particular, $T_W$ separates the high-temperature regime for which the Curie-Weiss law holds [S26]

$$\chi \simeq \frac{(g\mu_B)^2}{3}\frac{S(S+1)}{T-T_W},$$

from the low-temperature regime for which $\chi$ is strongly enhanced and given by Eq. (S9). The true critical temperature for this model is $T_C = 0$.

[S36] E. Strambini, V. N. Golovach, G. De Simoni, J. S. Moodera, F. S. Bergeret, and F. Giazotto, Phys. Rev. Materials **1**, 054402 (2017).

[S37] A. Brataas, Y. V. Nazarov, and G. E. W. Bauer, Phys. Rev. Lett. **84**, 2481 (2000).

[S38] Y.-T. Chen, et al., Phys. Rev. B **87**, 144411 (2013).